# An empirical assessment of best-answer prediction models in technical Q&A sites

Fabio Calefato[1]  · Filippo Lanubile[2] · Nicole Novielli[2]

## Abstract

Technical Q&A sites have become essential for software engineers as they constantly seek help from other experts to solve their work problems. Despite their success, many questions remain unresolved, sometimes because the asker does not acknowledge any helpful answer. In these cases, an information seeker can only browse all the answers within a question thread to assess their quality as potential solutions. We approach this time-consuming problem as a binary-classification task where a best-answer prediction model is built to identify the accepted answer among those within a resolved question thread, and the candidate solutions to those questions that have received answers but are still unresolved. In this paper, we report on a study aimed at assessing 26 best-answer prediction models in two steps. First, we study how models perform when predicting best answers in Stack Overflow, the most popular Q&A site for software engineers. Then, we assess performance in a cross-platform setting where the prediction models are trained on Stack Overflow and tested on other technical Q&A sites. Our findings show that the choice of the classifier and automatied parameter tuning have a large impact on the prediction of the best answer. We also demonstrate that our approach to the best-answer prediction problem is generalizable across technical Q&A sites. Finally, we provide practical recommendations to Q&A platform designers to curate and preserve the crowdsourced knowledge shared through these sites.

✉ Fabio Calefato
fabio.calefato@uniba.it

Filippo Lanubile
filippo.lanubile@uniba.it

Nicole Novielli
nicole.novielli@uniba.it

[1] Dipartimento Jonico, University of Bari "A. Moro", via Duomo, 259, -74123 Taranto, Italy
[2] Dipartimento di Infomatica, University of Bari "A. Moro", via E. Orabona, 4 -, 70125 Bari, Italy

## 1 Introduction

As projects grow in complexity, software developers seek information and technical support from experts outside their inner circles (Squire 2015). This trend has been steadily increasing over the years, at first through mailing lists, then through web-based discussion forums, leading in the last decade to the success of community-based, question-and-answer (Q&A) sites. Stack Overflow is the most popular technical Q&A site and represents the platform of choice when programmers need to ask for technical help.[1] Because of its impact, it is fundamental that this huge amount of crowdsourced knowledge remains well curated.

Technical Q&A sites have had a profound impact on the way programmers seek and provide knowledge (Vasilescu et al. 2014; Abdalkareem et al. 2017). To make the process of knowledge building sustainable, modern Q&A sites rely on gamification features, such as collaborative voting and badges, which motivate experts to quickly answer questions in exchange for visibility and reputation gain within the community (Mamykina et al. 2011). However, the very distinguishing feature of modern Q&A sites is the best answer annotation, that is, letting askers mark one answer as the accepted solution to their help request. This feature ensures that best answers surface to the top and become an asset for those seeking a solution to the same problem in the future (Anderson et al. 2012).

Despite its success, one of the greatest issues with Stack Overflow is *question starvation* (Nie et al. 2017), that is, the existence of question threads without an accepted solution. As of this writing, about 7 million questions (~50%) are still unresolved, albeit most of them (about 4.7 M, ~70%) are provided with answers. In such cases, software engineers need help to assess the quality of all the answers within a question thread to identify the potential solutions. We approach this problem as a binary-classification task where a *best-answer prediction* model is built to identify the *best answers* within question threads (Adamic et al. 2008). In real usage scenarios, for questions that have been resolved, the model would predict whether each answer has been marked or not as the actual accepted solution by the asker; instead, for unresolved questions, given the absence of an accepted solution, the same model would predict whether each answer is a potential solution – that is, one that might work for information seekers, even though it has not been marked as accepted. In fact, it is not uncommon for users, especially novices, to forget marking valid answers as a solution (Calefato et al. 2015).

Furthermore, Stack Overflow helps users assess the quality of answers to unresolved questions by sorting answers according to upvote and moving those with the largest scores to the top of the thread. However, one issue of vote-ordering is that the answers entered last, even when of high quality, tend to receive few or no upvotes and thus stay at the bottom of the thread; instead, the first ones keep receiving upvotes, benefiting from the Matthew effect of accumulated advantage (Mamykina et al. 2011; Roy et al. 2017) and, thus, remain on top. This issue of preferential attachment calls for the development of prediction models that suggest potential solutions to information seekers not exclusively based on the number of answer upvotes. Besides, given the sheer volume of posts already hosted at Stack Overflow and the growth rate of ~6 K new questions and ~8.5 K new answers posted per day over the last year,[2] we argue that models focusing on features that are computationally intensive have limited practical value.

---

[1] http://stackoverflow.com/research/developer-survey-2016
[2] http://data.stackexchange.com/stackoverflow/query/759408/daily-posts-in-2016

Because there is at most one best answer in a question thread, in any dataset, there are intrinsically fewer best answers (minority class) than non-accepted answers (majority class). This problem, which is referred to as *class imbalance* (He and Garcia 2009), is a fundamental issue of classification tasks because it can significantly compromise both the performance of learning algorithms and the reliability of assessment metrics. Besides, Japkowicz and Stephen (2002) observed that class imbalance is a 'relative problem' that depends on contextual factors such as the degree of imbalance, the size of the datasets, and the classifiers involved. Previous studies on best-answer prediction have overlooked the problem of learning from imbalanced datasets (Adamic et al. 2008; Burel et al. 2012; Shah and Pomerantz 2010; Tian et al. 2013). Furthermore, they do not encourage the cross-comparison of results as they apply different, sometimes inadequate, performance measures.

**Research Questions** In this paper, we aim to identify the most effective approach for predicting best answers in technical Q&A sites, while also focusing on selecting features that are computationally inexpensive. Given the imbalanced dataset problem, we review prior work on binary classification to select and use adequate metrics for a reliable assessment of the performance of best-answer prediction models.

To investigate the problem of best-answer prediction, we first focus on Stack Overflow, as it includes the largest amount of crowd-assessed best answers. Specifically, we choose Stack Overflow as the training set and we investigate to what extent classifiers can predict best answers when trained and tested on the Stack Overflow site. Thus, we formulate the first research question as follows:

*RQ1: How do classifiers perform for best-answer prediction within Stack Overflow?*

Stack Overflow, however, is not the only venue for requesting technical help. There are other Q&A platforms where developers ask for help about development in general (e.g., Programming & Design category in Yahoo! Answer),[3] or for specific products (e.g., SAP Community Network).[4] Furthermore, lots of crowdsourced knowledge is available in legacy developer-support tools, such as web forums and mailing lists, where developers typically used to seek for help before the advent of modern Q&A platforms. The benefit of best-answer prediction is even larger for web forums as they typically lack the features for marking an answer as accepted and moving it on top of the thread, thus potentially saving developers from the effort of looking into each answer of a thread to assess which solution is the one working. Accordingly, we investigate how classification models trained on Stack Overflow can predict best answers in other developer-support Q&A sites, including both legacy web forums and modern Q&A platforms. Web forums are included because of their relevance, since they lack the feature of marking best answers, whereas Q&A platforms are included for assessing to a larger extent the generalizability and robustness of the cross-prediction model. Therefore, we define our second research question as follows:

*RQ2: How do classifiers perform for cross-platform best-answer prediction?*

---

[3] https://answers.yahoo.com
[4] http://scn.sap.com

We build our methodology upon studies based on imbalanced datasets. Due to the limited amount of research on best-answer prediction, we also define the methodology based on previous work on software defect prediction, which studied prediction models in cross-context settings, that is, across projects and companies (Turhan et al. 2009; Zimmermann et al. 2009). Then, we conduct a study in two stages, one for each of the two research questions. All the datasets and scripts used in this study are made publicly available.[5]

From the results of our study, we observe that:

1. There is a clear difference in performance among classification techniques, with performance improvements by over 40 percentage points in terms of AUC – i.e., the choice of the classifier and its parameter configuration has a significant and large impact on the performance of the best-answer prediction model to build.
2. The performance of models in the cross-platform setting is above the trivial rejector baseline and is comparable to the upper-bound performance achieved by training the same prediction models in a within-platform setting on all the test sets – i.e., our approach to the best-answer prediction problem is generalizable across technical Q&A sites.

**Contributions** This study extends our earlier work (Calefato et al. 2016) in which we assessed only 4 classifiers, all from the family of decisions trees. Furthermore, in this work, we provide a more robust empirical evaluation including automated parameter tuning, feature selection, and, finally, employing performance metrics adequate in case of class imbalance. A more detailed comparison is available in Section 7 – Related work.

From a research perspective, this is – to the best of our knowledge – the first study on best-answer prediction that (i) analyzes the performance of multiple classifiers in a cross-context setting, where models are trained and tested on corpora from different technical Q&A platforms, and (ii) uses a systematic approach to deal with the problem of class imbalance.

Because we rely on training and test sets that come from entirely distinct data sources, our findings gain in robustness and generalizability. As such, another research contribution is the identification of computationally inexpensive features that are essential for best-answer prediction in both modern and legacy platforms, rather than accidental because specific of one technical Q&A site. Besides, most prior studies (i) built models using one classification technique only, (ii) failed to visually assess differences by plotting performance curves, and (iii) reported performance by single scalar measures that are unstable and sensitive to dataset imbalance. These limitations not only prevent a proper comparison with our study findings but, more in general, they also reduce the possibility of replicating experiments on best-answer prediction. Therefore, as a further research contribution, in this paper, we build a reliable benchmark for best-answer prediction studies, with recommended measures and data from five different datasets, which sets an initial 'baseline' for the evaluation of future research efforts.

From a practical perspective, this research furthers the state of practice by providing solutions to some recurring problems such as (i) the numerous question threads that remain unresolved in technical Q&A sites, (ii) ensuring the quality of the crowdsourced knowledge stored therein, and (iii) the loss of knowledge caused by the phenomenon of developer-support communities migrating from legacy forums to modern Q&A sites. In fact, as of this writing, the entire Stack Overflow site hosts over 14 million questions of which only about 7 million are provided with an accepted answer. Those unresolved, however, are not just 'stale' questions asked many years ago: almost 4 million questions asked on Stack Overflow in the last 2.5 years are still left open. Sometimes, questions are not closed because no good answer was provided; other times, it is just because the asker did not take any action, albeit in presence of useful answers. Yet, in every thread, there are some answers that are better than the others. Finding good answers is even worse when threads are popular because it forces developers seeking a solution to go through all the responses just relying on the number of upvotes as a proxy measure of their quality. Q&A sites might implement a recommender system that leverages our approach to highlight potential solutions when users browse unresolved questions without an accepted solution, as a preliminary step towards identifying and marking the resolving answer. Besides, because the used features are not computationally expensive, recommendations could be computed and updated on the fly, ensuring that also high-quality, newly-entered answers have fair chances to be recommended against the first and most upvoted ones. More in general, by implementing our solution, Q&A platforms would ensure that unresolved questions become a more useful source of information to developers seeking for technical help.

---

[5] https://github.com/collab-uniba/emse_best-answer-prediction

Another practical contribution is related to the importance of ensuring that the huge amount of crowdsourced knowledge stored in Q&A sites is well curated. In fact, software developers typically need knowledge that goes beyond what they already possess and, therefore, they must seek further information (Shaw 2016). Anderson et al. (2012) have observed a shift in the purpose of these Q&A sites. While initially aimed at quickly providing solutions to information seekers' needs, modern Q&A sites have now become platforms supporting the creative process of community-driven knowledge with a long-term value to a broader audience. For example, Parnin et al. (2012) observed that questions and answers contributed on Stack Overflow covered about the 87% of the Android API classes. As such, it is fundamental to information seekers that this huge amount of crowdsourced knowledge available through modern Q&A sites is well curated – also from an educational perspective, considering that most of the software developers using Stack Overflow (~70%) report to be at least partly self-taught (Stack Overflow Developer Survey 2016).

Finally, a side effect of the Stack Overflow popularity is the growing trend of developer communities that are abandoning their legacy support forums over Stack Overflow, seeking more visibility, larger audience, and faster response times. Despite the evident benefits of moving to a modern and very successful infrastructure, one consequent downside is that the history and the crowdsourced knowledge generated through these legacy forums is at stake of being lost. For instance, based on the analysis of 21 developer support communities that moved from legacy sites to Stack Overflow between 2011 and 2014 (Calefato et al. 2016), we found that about 20% of them had not archived their content, thus causing a loss of knowledge that still surfaced in Google searches but turned out to be actually inaccessible. Moving the content of a legacy forum to a modern Q&A site would not only preserve all that crowdsourced knowledge but also increase its long-term value. Our research has the potential of assisting in the migration of the content from a legacy developer forum to a modern Q&A platform because the former typically lacks the feature of marking an answer in a thread as the accepted solution. Therefore, our research can help identify the candidate solutions in the entire history of question threads to be imported into the modern Q&A site of choice. To date, none of the available modern Q&A platforms allows importing existing content from other sources. Also, the migration of content comes with other challenges, such as coping with different community cultures and user identities. Some initial work has been performed by Vasilescu et al. (2014) who investigated how to match the identities of the R-help community members after the migration onto the Stack Overflow and Cross Validated Q&A sites from the original mailing list. Here, instead, we demonstrate the technical feasibility of content migration from legacy to modern Q&A platforms. Finally, our research has also the potential to help companies that want to run a private version of Stack Overflow[6] without having to lose the existing base of knowledge or maintain it elsewhere.

**Structure of the Paper** The remainder of this paper is organized as follows. In Section 2, we outline the methodology for this study. In Sections 3 and 4, respectively, we illustrate the datasets and the features extracted. Then, we illustrate the study settings and the findings, respectively in Section 5 and 6. In Section 7, we review previous work on best-answer prediction and discuss our findings. Study limitations are presented in Section 8 and conclusions are drawn in Section 9.

## 2 Methodology

In this section, we define the methodology that we followed to perform the empirical study on best-answer prediction. Our goal is to identify the adequate solutions and metrics to assess the performance of best-answer prediction models, trained and tested on imbalanced datasets from different Q&A platforms.

### 2.1 Prediction Models

The problem of using prediction models to identify best answers in technical Q&A sites is relatively new. Nonetheless, research in software engineering has already been using prediction models for decades, especially to identify defective code (Arisholm et al. 2010; Catal and Diri 2009). To do so, historical project data about known bugs are used to train binary classification models that are then used to classify unseen software modules as defective or not (Menzies et al. 2010).

Because most of the standard learning algorithms assume a balanced training set, imbalanced datasets may generate suboptimal classification models that provide a good coverage of the majority examples, whereas the minority ones are frequently misclassified if not completely discarded as noise (He and Garcia 2009; Lopez et al. 2013). Besides, learning algorithms are often built to maximize metrics such as *Accuracy*, which is not the most appropriate performance measure for imbalanced problems (see Section 2.3) (Lemnaru and Potolea 2011). In recent years, ensembles of classifiers have arisen as a possible solution to the class imbalance problem (Lopez et al. 2013). Ensemble learners, also known as multiple classifier systems, try to improve the performance of single classifiers by inducing several classifiers and combining them to obtain a new classifier that outperforms every one of them (Polikar 2006). These approaches have shown to be very accurate. The downsides are the increased learning time and difficulty in comprehending the output model.

The techniques reported in Table 1 include the most-used classifiers for software defect prediction according to a systematic review by Malhotra (2015). In another recent literature review, Wahono (2015) found that these classifiers account for over the 75% of the identified primary studies. Finally, the 8 classification techniques listed in Table 1 mostly cover those investigated by Lessmann et al. (2008) in their seminal study for benchmarking defect prediction classifiers.

---
[6] https://business.stackoverflow.com/enterprise

**Table 1** A breakdown of the most used classifiers and techniques employed in defect prediction studies

| Technique | Classifier |
|---|---|
| Regression-based | Logistic Regression |
| Bayesian | Naïve Bayes |
| Nearest Neighbors | K-Nearest Neighbors |
| Decision Trees | C4.5 / J48 |
| Support Vector Machines | Sequential Minimal Optimization |
| Neural Networks | Radial Basis Functions |
| Ensemble (Bagging) | Random Forests |
| Ensemble (Boosting) | Adaptive Boosting |

In this study, we assess 26 classifiers overall. To select them, we started from the 8 most used classifiers reported in Table 1. Then, we extended the set with the classifiers employed for software defect prediction in a recent study by Tantithamthavorn et al. (2016), who studied the effects of leaners selection on classification performance – a scenario similar to our intended assessment of best-answer prediction models. The complete list of the classifiers employed in this study is given later in Section 5 (see Table 7).

Because performance differences can be large, previous research recommends the inclusion of a preliminary assessment of various classification algorithms, as a former step before selecting the most appropriate for the given context (D'Ambros et al. 2012; Denaro and Pezzè 2002; Ghotra et al. 2015). However, no general consensus exists about any specific framework to apply for the assessment of model performance (Demsar 2006; Lessmann et al. 2008; Jiang et al. 2008a; Zhang and Zhang 2007). Recently, Ghotra et al. (2015) have suggested an analysis procedure based on the Scott-Knott test (Scott and Knott 1974) to find statistically distinct clusters of classification techniques. The Scott-Knott test is a mean-comparison approach that uses hierarchical cluster analysis to recursively partition a set of treatment means (e.g., means of model performance) into groups until they can no longer be split into further, statistically-distinct groups. This test overcomes the confounding issue of overlapping groups affecting other clustering algorithms such as the Nemenyi post hoc test (Ghotra et al. 2015). By overlapping we mean the possibility for one or more treatments to be classified in more than one group. Furthermore, the Scott-Knott test is also capable of coping with normal and non-normal distributions (Borges and Ferreira 2003). Recently, Tantithamthavorn et al. (2017) have recommended using the Scott-Knott ESD (Effect Size Difference), an enhancement over the standard Scott-Knott test, which uses Cohen's effect size to partition groups and ensure that (i) the magnitude of the difference for all of the treatments in each group is negligible and (ii) the magnitude of the difference of treatments between groups is non-negligible.

> *We apply the Scott-Knott ESD test to compare the performance of 26 classifiers.*

For projects that are new or just lack historical data on which to build defect-prediction models, researchers have proposed *cross-project defect prediction*, that is, reusing models built on datasets from other projects (Nam and Kim 2015). Previous research has shown that, albeit viable (Kitchenham et al. 2007), cross-project defect prediction is a real challenge, especially if conducted on projects from different companies (Turhan et al. 2009). In particular, the conditions under which it is effective are still not entirely clear (Kitchenham et al. 2007). In their study on over 600 cross-project predictions, Zimmermann et al. (2009) observed that simply using datasets from projects within the same domain does not always work to build effective models. Instead, they observed that prediction models perform better when (i) datasets from projects with similar characteristics are used (e.g., both open source, same size and domain, etc.) and (ii) they are trained on a project dataset that is larger than the target.

The two research questions defined are already in line with the recommendations above. As further illustrated in Section 3, we use Stack Overflow (i.e., the largest Q&A site) as the training set for building a best-answer prediction model that is fit to the data collected from different technical Q&A platforms, both modern and legacy.

## 2.2 Parameter tuning and feature selection

Tuning parameters in classifiers is important because it changes the heuristics determining how they learn. For example, it controls the number of decision trees to use in Random Forest or the number of clusters in $K$-Nearest Neighbors (KNN). Research on software defect prediction has widely acknowledged that models trained with suboptimal parameter settings may underperform as parameter settings depend on the dataset (Hall et al. 2012; Jiang et al. 2008b; Mende and Koschke 2009; Tosun and Bener 2009). Furthermore, Ghotra et al. (2015) found that without parameter tuning, most classification techniques produce models whose performances are statistically indistinguishable. Nonetheless, parameters are often left at their default values, even though the implementations of the classifiers provided by data mining toolkits such as R,[7] Weka,[8] and scikit-learn,[9] typically use very different default settings. Therefore, relying on default parameter is a common issue that severely limits the replicability of classification experiments (Menzies and Shepperd 2012).

Manual tuning has been so far impractical (Bergstra and Bengio 2012). For example, Kocaguneli et al. (2012) found that there are about 17,000 possible setting combinations to explore for the KNN classifier. However, Fu et al. (2016) have recently developed a technique for automated parameter tuning, which avoids the need to explore all the possible combinations. They found that parameter tuning (i) only takes a low number of evaluations (usually tens) to identify the optimal setting; (ii) it improves predicting performance up to 60%; (iii) it changes what features are the most important to the prediction; and (iv) it is heavily dataset-dependent. These findings have been confirmed by Tantithamthavorn et al. (2016), who performed a similar study of the effect of automated parameter tuning for a couple of dozens of classifiers. In addition, they observed an increase in stability of classification performance, especially for cross-project defect prediction.

> *We perform automated parameter tuning to optimize performance when building best-answer prediction models.*

Fu et al. (2016) have also observed that different subsets of relevant features are selected when parameters are tuned. This is because parameter tuning changes how classification algorithms explore the space of possible models; therefore, if the exploration process is modified, so are the features found by it. Among the techniques available for feature selection, Menzies (2016) recommends wrapper-based feature selection methods. Wrapper methods consider feature selection as a search problem where multiple models are evaluated using procedures that add and/or remove predictors to find the optimal combination that maximizes model performance. In essence, wrapper methods are search algorithms that treat predictors as the inputs and model performance as the output to be optimized (Karegowda et al. 2010). Albeit wrapper methods generally achieve good performance, they are also expensive for large datasets in terms of computational complexity and time since each feature combination must be evaluated with the algorithm used.

---

[7] https://www.r-project.org  
[8] http://www.cs.waikato.ac.nz/ml/weka  
[9] http://scikit-learn.org

> *We perform wrapper-based feature selection to identify the set of most relevant features for best-answer prediction.*

### 2.3 Evaluation criteria

When learning from imbalanced datasets, the evaluation of classification performance must be carried out using specific metrics to consider the class distribution and properly assess the effectiveness of learning algorithms. Therefore, the selection of the evaluation criteria here is a key factor for a reliable assessment of the classification performance of prediction models.

In a binary classification problem like best-answer prediction, a confusion matrix (or contingency table, see Table 2) records the results of correctly and incorrectly recognized examples of each class. From a confusion matrix, we can obtain several metrics (see Table 3) for the classification performance of both positive and negative classes independently. In the following, we review several performance metrics and select the most adequate to our study setting.

Traditionally, the machine learning community has used the scalar measures of prediction *Accuracy* ($PA$) and *Error rate* ($E = 1 - PA$) as simple performance metrics for binary classification problems (see Table 3). However, neither *Accuracy* nor *Error rate* is able to provide adequate information about a classifier performance in the framework of imbalanced datasets (Provost et al. 1998; Ringrose and Hand 1997). In fact, due to the way they are defined, they cannot distinguish between the number of correctly classified examples of different classes (Yang and Liu 1999). As such, being highly sensitive to changes in data, they may lead to erroneous conclusions (He and Garcia 2009). For example, a rejector that classifies all instances as negatives may paradoxically achieve an accuracy of 90% in a dataset with a ratio of 1:9 between the two classes in the dataset, and possibly outperform all nontrivial classifiers.

*Precision* is better able to capture the effect on a classifier performance of having a larger number of negative examples (Davis and Goadrich 2006). Nonetheless, *Precision* is still sensitive to changes in data distribution and it cannot assert how many positive examples are classified incorrectly (He and Garcia 2009). Furthermore, Menzies et al. (2007) found that

**Table 2** Confusion matrix for a two-class problem

|　|　| Prediction | |
|---|---|---|---|
|　|　| Positive | Negative |
| Actual | Positive | True Positives (TP) | False Negatives (FN) |
|　| Negative | False Positives (FP) | True Negatives (TN) |

Table 3 Performance metrics for binary classification problems obtained from confusion matrix

| Metrics (synonyms) | Definition | Description |
| --- | --- | --- |
| Accuracy | $Acc = \frac{TP+TN}{TP+FN+FP+TN}$ | Proportion of correctly classified instances |
| Error rate | $E = 1 - Acc$ | Proportion of incorrectly classified instances |
| Precision (*Positive Predicted Values*) | $P = \frac{TP}{TP+FP}$ | Proportion of instances correctly classified as positive |
| Recall (*Probability of Detection, True Positive rate, Sensitivity*) | $R = TP_{rate} = \frac{TP}{TP+FN}$ | Proportion of positive instances correctly classified |
| F-measure (*F1-score*) | $F = 2\frac{P \times R}{P+R}$ | Harmonic mean of Precision and Recall |
| True Negative rate (*Specificity*) | $TN_{rate} = \frac{TN}{TN+FP}$ | Proportion of negative instances correctly classified |
| G-mean | $G = \sqrt{TP_{rate} \times TN_{rate}}$ | Geometric mean of True Positive rate and True Negative rate |
| False Positive rate (*Probability of False Alarm*) | $FP_{rate} = \frac{FP}{FP+TN}$ | Proportion of negative instances misclassified |
| AUC (*AUCROC*) | $AUC = \int_{-\infty}^{+\infty} R(T)\left(-FP'_{rate}(T)\right) dT$ | The area under the graphical plot of Recall vs. $FP_{rate}$ for a binary classifier as its discrimination, cutoff threshold is varied |
| Balance | $Bal = 1 - \frac{\sqrt{(0-FP_{rate})^2 + (1-R)^2}}{\sqrt{2}}$ | Euclidean distance from the point ($FP_{rate}$, $R$) of the ROC curve from the point ($FP_{rate} = 0$, $R = 1$), representing the perfect classifier in the ROC space |

Positive class = TP + FN and Negative class = FP + TN

*Precision* has instability issues, i.e., large standard deviations when coping with skewed class distributions, which make it very difficult to show that a classifier outperforms another in its sole terms. Unlike *Precision*, *Recall* is not sensitive to data distribution. However, any assessment based solely on *Recall* would be inadequate, as it provides no insight of how many examples are incorrectly classified as positives.

None of these scalar metrics can provide a reliable assessment of classification performance when used by itself. Nonetheless, these individual scalar metrics can be combined to build more reliable measures of classification performance, especially in imbalanced scenarios (Liu et al. 2006). Specifically, these aggregated metrics of performance (see Table 3) include *F-measure* – i.e., the harmonic mean of *Precision* and *Recall*, and *G-mean* – i.e., the geometric mean between *True Positive rate* ($TP_{rate}$, also known as *Recall*) and *True Negative rate* ($TN_{rate}$, the proportion of negative instances correctly classified). However, prior research (e.g., (Rahman et al. 2012)) has reported that these scalar metrics are all threshold dependent, that is, they depend on the 'cutoff' value used to decide whether an instance is to be classified as either positive or negative when computing the confusion matrix. In other words, a threshold probability is selected (by default .50), above which predictions are classified as positive and below which they are classified as negative. The default threshold works well when the instances in the two classes are equally distributed, however, this is not the case with imbalanced datasets.

Unlike scalar measures that impose a one-dimensional ordering, two-dimensional plots preserve all the performance-related information about a classifier, thus allowing for a visual analysis and comparison of the results from binary classifications (Drummond and Holte 2006). The *Receiver Operating Characteristic* (ROC) plot (Provost and Fawcett 1997) is a two-dimensional graphic with $FP_{rate}$ as the x-axis and the $TP_{rate}$ as the y-axis (see Fig. 1). A ROC plot allows the performance visualization of a binary classifier as the trade-off between the accurate classification of the positive instances and the misclassification of the negative instances (Hastie et al. 2009). A curve in the ROC space is formed from a set of points obtained by choosing different 'cutoffs' (thresholds) to assign probabilistic predictions to either class. The points (0, 0) and (Adamic et al. 2008) represent the trivial classifiers that classify all the instances, respectively, as negative and positive. The diagonal line connecting these two points represents random performance. In the ROC space, the goal is to be in the upper-left-hand corner, as close as possible to the point (0, 1) representing the perfect classification performance. As such, when comparing the performance of, say, two classifiers, one is said to 'dominate' the other when the ROC curve of the former is above and to the left of the other's – i.e., it has better accuracy on both positive and negative class. According to Fawcett (2006), ROC curves are not sensitive to changes in the skewness of class distribution, thus making them particularly suited to compare classifiers over different datasets.

The *Area Under the ROC Curve* (AUC or AUROC) is a scalar value representing the ROC performance of a classifier (see Table 3). AUC has an important statistical property: it represents the probability that a classifier will rank a randomly chosen positive instance higher than a randomly chosen negative one – i.e., the larger the area under the ROC curve, the higher the classification potential of a classifier (Fawcett 2006). An AUC value is between 0 and 1. However, because random guessing produces the diagonal line between (0, 0) and (Adamic et al. 2008), which has an area of 0.5, no realistic classifier should have an AUC less than 0.5. Since it is threshold independent, AUC provides a reliable measure to compare the performance of different classifiers (Lessmann et al. 2008) even with imbalanced datasets (Huang and Ling 2005). Furthermore, Adams and Hand (1999) recommend reporting AUC values with ROC plots. Similar to AUC, *Balance* is another scalar measure related to the ROC plot (see Table 3). *Balance*, in fact, measures the Euclidean distance from the point (0, 1) representing the perfect classifier in the ROC space (Hall et al. 2012).

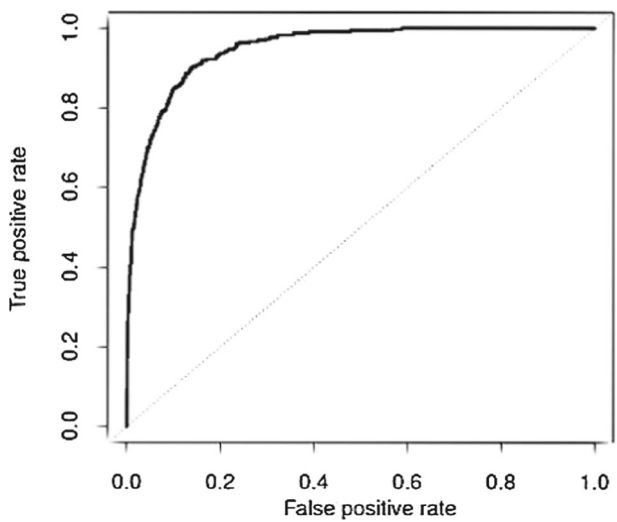

**Fig. 1** ROC plot for graphical assessment of performance

Rahman et al. (2012) and Zhang et al. (2016) observed that threshold-insensitive metrics are better suited to assess the performance in cross-prediction contexts since they do not require choosing the optimal cut-off value as other scalar metrics.

*We select AUC and Balance as scalar metrics representing the ROC performance of best-answer prediction.*
*We draw ROC plots to perform a graphic comparison of the performance of multiple classifiers.*

## 3 Datasets

We use five datasets, one as training set and the remaining four as test sets. Because all the datasets are imbalanced, we report the skewness of their class distribution using the *pos/neg ratio* (or *imbalance ratio*), a descriptive metric defined as the ratio of the number of instances in the positive (minority) class to the number of examples in the negative (majority) class (Lopez et al. 2013).

In the remainder of this section, we first present the training set built from Stack Overflow. Then, we present the four test sets, distinguishing between legacy developer-support forums and modern Q&A platforms; such different test sets allow us to investigate the extent to which a model trained on Stack Overflow can predict best answers across different platforms.

### 3.1 Training set: Stack overflow

The training set was built using the official dump of Stack Overflow[10] released in February 2016. We randomly selected a subset of questions threads from the 40 M questions in the dump. Therefore, the final training set contains about 507 K questions and 1.37 M answers. We use the accepted answers (~279 K) as positive examples to train the best-answer classifiers, whereas the remaining unaccepted answers (~1.09 M) represent the negative examples. As shown in Table 4, the Stack Overflow training set is skewed towards unaccepted answers, with a pos/neg ratio of about 1:4.

### 3.2 Test sets: Legacy forums

The first test dataset is *DocuSign*,[11] an electronic-signature API for securing digital transactions. The original support forum of DocuSign has been abandoned as its developer community moved to Stack Overflow. In June 2013, the forum became read-only, inviting community members to use Stack Overflow with the custom tag docusign-api. By the end of 2015, the forum and all its content became completely unavailable.

The second test dataset was obtained from the developer-support forum of *Dwolla*,[12] a Web API for real-time digital payments. The forum was abandoned during mid-2014 but the content is still available on the web as of this writing.

---
[10] https://archive.org/details/stackexchange
[11] https://www.docusign.com
[12] https://www.dwolla.com

**Table 4** Breakdown of the datasets

| Dataset | | Q&A Platform | Questions threads | Questions resolved (%) | Answers | Answers accepted (%) | pos/neg ratio |
|---|---|---|---|---|---|---|---|
| Training | Stack Overflow (SO) | Modern | 507 K | 279 K (~55%) | 1.37 M | 279 K (~20%) | ~1:4 |
| Test | DocuSign (DS) | Legacy | 1572 | 473 (~30%) | 4750 | 473 (~10%) | ~1:10 |
| | Dwolla (DW) | Legacy | 103 | 50 (~48%) | 375 | 50 (~13%) | ~1:7 |
| | Yahoo! Answer (YA)* | Modern | 41,190 | 29,021 (~70%) | 104,746 | 29,021 (~28%) | ~1:4 |
| | SAP Community Net. (SCN)** | Modern | 35,544 | 9722 (~27%) | 141,692 | 9722 (~6%) | ~1:15 |

*Extracted category: Programming & Design

**Extracted spaces: Hana, Mobility, NetWeaver, Cloud, OLTP, Streaming, Analytic, PowerBuilder, Cross, 3d, Frontend, ABAP

To obtain a dump of the content from these two legacy forums, we built two custom scrapers using the Python library Scrapy.[13] The scrapers downloaded all the question threads from the two forums and stored them in a local database for convenience. Unlike other legacy developer-support sites but analogously to modern Q&A sites, the DocuSign original forum allowed the question asker to select one answer in the thread as the accepted solution. As such, the DocuSign dataset was already annotated with accepted answers. Instead, the Dwolla forum did not offer such feature and then a 'gold standard' was created by manually inspecting each extracted question thread in the dump to identify the accepted solution (if any) among the existing answers. Two researchers first performed the annotation independently; then, they compared the results and iteratively resolved conflicts until complete agreement was reached.

As shown in Table 4, the DocuSign dataset, which contains 4750 answers to 1572 questions threads, is skewed towards unaccepted answers since only the ~10% of questions are marked as accepted solutions, with a pos/neg imbalance ratio of about 1:10. Likewise, the Dwolla dataset has a similar pos/neg ratio of 1:7, albeit the size is quite smaller than DocuSign, with only 375 answers to 103 questions.

### 3.3 Test sets: Modern platforms

The modern Q&A sites from which the other two test sets were retrieved are Yahoo! Answers and SAP Community Network. Unlike the legacy forums that provide all the content mostly via pure HTML, these modern sites rely heavily on JavaScript to make their content available. Therefore, to obtain the dump of these two sites, we first loaded pages in Selenium,[14] a web driver capable of controlling the navigation of web pages, and then we scraped their content using Scrapy.

*Yahoo! Answer* is a modern community-driven Q&A site, launched in 2005 by Yahoo!, which relies on gamification to foster participation and quality of answers. Unlike Stack Overflow, however, Yahoo! Answers is a general-purpose Q&A site, organized by categories that range from politics to art. To ignore non-technical content, we retrieved only the technical help requests from the *Programming & Design* category.

---

[13] http://scrapy.org
[14] http://www.seleniumhq.org/projects/webdriver

*SAP Community Network* (SCN) is the official developer-support community of SAP AG. Released in 2003, SCN is organized into subgroups, called 'spaces', dedicated to specific SAP products, technologies, and programming languages (e.g., ABAP). In 2013, the forum was updated to support gamification and encourage community engagement on SCN. Although the SCN forum is still active, community members also post questions on Stack Overflow, using dedicated tags such as hana or abap. Therefore, we decided to retrieve content only from those SCN spaces for which a homonymous tag exist in Stack Overflow. Besides, unlike the other Q&A sites included in this study, SCN is the only platform supporting more than two states for a question, namely *open* (or unanswered), *supposed answered,* and *answered*. We aggregate the first two categories together because an assumed-answered question is used by a question asker who wants to close a discussion regardless of the lack of an accepted solution.

Although the two datasets are comparable in size, as shown in Table 4, the number of accepted answers in the Yahoo! Answer dataset is much larger than SCN, with about 29 K answers (~28%) marked as accepted in the former as opposed to only 9.7 K accepted answers (~6%) in the latter. Consequently, the Yahoo! Answer dataset has a pos/neg ratio of about 1:4, as opposed to SCN, which exhibits the largest imbalance ratio of 1:15 among the four test sets.

## 4 Features

In this section, we describe the features used by the classifiers to learn the prediction models.

The five datasets in this study come from data sources with different formats and structures. To facilitate their representation and analysis, we first converted their content into a common format and then stored it into a local database. Table 5 provides an overview of which relevant elements are available from each data source. The extracted information elements relate to either the *thread content* (e.g., question body and title) or the *thread metadata* (e.g., when an answer was entered, its number of views and upvotes received).

Crossed out elements are discarded, whereas the others are retained. Specifically, we discard all the information elements that are not available in Stack Overflow, since we use it

**Table 5** Relevant information elements extracted from the five dumps

| Information Elements | Stack Overflow | DocuSign | Dwolla | Yahoo! Answers | SCN |
|---|---|---|---|---|---|
| Type (question/answer) | Yes | Yes | Yes | Yes | Yes |
| Body | Yes | Yes | Yes | Yes | Yes |
| Title | Yes | Yes | Yes | Yes | Yes |
| ~~Answer tags~~ | No | Yes | No | No | No |
| URL | Yes | Yes | Yes | Yes | Yes |
| Question ID | Yes | Yes | Yes | Yes | Yes |
| Question closed | Yes | Yes | Yes | Yes | Yes |
| Answer count | Yes | Yes | Yes | Yes | Yes |
| Accepted answer ID | Yes | Yes | Yes* | Yes | Yes |
| Date/time | Yes | Yes | Yes | Yes | Yes |
| ~~Answer views~~ | No | Yes | No | No | No |
| Rating score (upvotes − downvotes) | Yes | Yes | No | Yes | Yes |

*After manual annotation

to train the prediction models in our study. Regarding the thread content, we observe that the element *answer tags* is only available for the DocuSign test set and, therefore, it is removed. Likewise, with respect to the thread metadata, we discard the *answer views* information element because it is only available from DocuSign.

All the retained information elements in Table 5 provide a set of common features available in all the five datasets. Besides, we also exclude any information element related to user profiles and gamification (e.g., reputation, badges, the number of accepted answers and favorited posts). This is because they are generally not available in legacy developer-support forums and, even when existing, user-related features are very dynamic in nature and, thus, they need to be constantly recomputed.

The information elements listed in Table 5 are used to define the set of features employed to train and test the prediction models. The overall set of 22 features is reported in Table 6, arranged in four main categories: *linguistic*, *vocabulary*, *meta*, and *thread*.

*Linguistic* features represent the attributes of questions and answers and are intended to estimate their quality. Such *linguistic* features are called 'shallow' because they measure readability through the 'surface' properties of a text, such as the number of words and the average word length (Pitler and Nenkova 2008). As such, they are also computationally cheap. For example, the *linguistic* features include *length* (in characters), *word count*, and *average number of words per sentence* in an answer. We also add a boolean feature to this category, namely *contains hyperlinks*, because in our previous work we found that the presence of links to external resources is positively related to the perception of answer completeness (Calefato et al. 2015).

To estimate the readability of an answer we also employ two *vocabulary* features, i.e., *normalized Log Likelihood* and *Flesch-Kincaid grade*. The *normalized Log Likelihood* ($LL_n$), already employed in previous work (Gkotsis et al. 2014; Gkotsis et al. 2015; Pitler and Nenkova 2008) uses a probabilistic approach to measure to what extent the lexicon in an answer is distant from the vocabulary used in the whole forum community. Specifically, $LL_n$ is defined as follows:

$$LL_n = \frac{LL = \sum^{w_s} C(w_s) \log(P(w_s|Voc))}{UC(w_s)} \quad (1)$$

Table 6 Summary of the overall 22 features in our model, arranged by category

| Category | Feature | Ranked version |
| --- | --- | --- |
| Linguistic | Length (in characters) | Length_ranked |
|  | Word count | Word count_ranked |
|  | No. of sentences | No. of sentences_ranked |
|  | Longest sentence (in characters) | Longest sentence_ranked |
|  | Avg words per sentence | Avg. words per sent._ranked |
|  | Avg chars per word | Avg. chars per word_ranked |
|  | Contains hyperlinks (boolean) | – |
| Vocabulary | $LL_n$ | $LL_n$_ranked |
|  | F-K | F-K_ranked |
| Meta | Age (hh:mm:ss) | Age_ranked |
|  | Rating score (upvotes – downvotes) | Rating score_ranked |
| Thread | Answer count | – |

Given $s$, a sentence in an answer, $P(ws|Voc)$ is the probability of the word $ws$ to occur, according to the background corpus $Voc$, and $C(ws)$ is the number of times the word $ws$ occurs in $s$. $LL$ is normalized by dividing it by the number of unique words occurring in $s$. The normalization in (Adamic et al. 2008) is necessary to consider answers of different lengths. $LL_n$ is the most computationally intensive feature in our set.

The *Flesch-Kincaid grade* (*F-K*) is another readability metric defined by Kincaid et al. (1975) and already used by Burel et al. (2012) in previous work. *F-K* is calculated as follows:

$$F\text{-}K_{p_i}\left(awsp_{p_i}, asps_{p_i}\right) = 0.39\ awsp_{p_i} + 11.8\ asps_{p_i} - 15.59 \quad (2)$$

In the definition (Adams and Hand 1999) above, for any given post $pi$, $awps_{pi}$ is the average number of words per sentence and $asps_{pi}$ is the number of syllables per word.

Two are the features belonging to the *meta* category. The first one, *age*, applies only to answers and it computes the time difference since the question has been posted. The time dimension is a predictor that cannot be overlooked because the best (accepted) answer is often the fastest among a set of good-enough answers (Treude et al. 2011). The second one, *rating score*, is the number of upvotes minus the number of downvotes that an answer has received from community members. This feature reflects the perceived usefulness of an answer.

The *thread* category includes only one feature, *answer count*, that is, the number of answers to a question. As such, this feature reflects the popularity of a thread.

Furthermore, as shown in the rightmost column of Table 6, for all the features, except *answer count* and *contains hyperlinks*, we also assign ranks after computing their numeric values. In other words, for each question thread, we group all answers, compute a feature, and then rank the values in either ascending or descending order. In the following, we refer to this procedure as *ranking* (Calefato et al. 2016). For instance, for the *word count* linguistic feature, the answer in the thread with the largest value ranks 1, the second largest ranks 2, and so on (i.e., descending order) because we assume that long answers are more accurate and elaborate and, hence, have a larger chance of being accepted. For the *age* feature, instead, we assign the rank 1 to the quickest answer (i.e., ascending order), because previous research has demonstrated that responsiveness plays a major role in getting answers accepted in Q&A websites (Mamykina et al. 2011; Wang et al. 2017).

## 5 Empirical study

Our assessment framework is divided into two stages, one per research question: (Adamic et al. 2008) best-answer prediction within Stack Overflow and (Adams and Hand 1999) cross-platform best-answer prediction. Both stages are carried out using the R statistical software package.

### 5.1 Best-answer prediction within stack overflow

As this stage is concerned only with the Stack Overflow dataset, the other four test sets are left out and used in the next assessment stage.

Given the size of the Stack Overflow dataset (~1.4 M answers), to reduce the computational time, we generated a subsample of about 341 K answers with the same pos/neg ratio as the whole dataset. Model evaluation starts by seeking the optimal parameter configuration for the 26 classifiers (Fig. 2). Assessing all the possible settings in the parameter spaces would be unfeasible. Thus, we rely on the caret[15] R package to perform automated parameter optimization in three steps. In the first step, caret generates the candidate parameters for each of the classifiers. We set the tuneLength parameter to 5, as suggested by Kuhn (2008) and Tantithamthavorn et al. (2016); this parameter represents a 'budget threshold' that constrains the maximum number of different values to be evaluated for each parameter. Then, in the second step, caret evaluates all the potential combination of the candidate parameter settings until the budget threshold is exceeded. Finally, in the third step, caret outputs the optimized settings that achieve the highest performance estimate. Only 3 out of 26 classifiers (see Table 7) require no tuning as they do not have parameters (i.e., GLM, LDA, and BaggedCART).

At this stage, the best-answer prediction models are evaluated using a 10-fold cross-validation approach, which divides the input dataset into 10 folds of equal size. Of the 10 folds, 9 are allocated to the training corpus (90%) and 1 is set aside for the testing (10%). This process is repeated ten times, using each fold as the testing fold once. To further increase the validity of our results, we repeat the entire 10-fold process ten times, for a total of 100 iterations; through these iterations, we reduce the amount of variance in the evaluation of the prediction models. Despite the class imbalance in the Stack Overflow dataset, as suggested by Turhan (2012), we do not apply any resample technique to rebalance the dataset. Instead, we use a stratified technique to generate folds that preserve in both the training and test sets the same class distribution as in the original dataset.

To compare the performance of the best-answer prediction models, as suggested by Tantithamthavorn et al. (2017), we use the Scott-Knott ESD test, which groups the models into statistically distinct clusters with a non-negligible difference, at level $\alpha = 0.01$. The grouping is performed based on mean AUC values (i.e., the mean AUC value of the ten × 10-fold runs for each prediction model). Therefore, this evaluation stage eventually results in a list of all the studied classifiers ranked by performance (mean AUC), from which we select the top-ranked ones that are statistically better at the 1% level. In addition, to graphically identify further differences in performance, we also plot the ROC curves for select models.

### 5.2 Cross-platform best-answer prediction

The second stage concerns with evaluating the performance of the selected top-ranked classifiers in a cross-platform setting, i.e., trained on Stack Overflow and tested on four test sets, namely DocuSign, Dwolla, Yahoo! Answers, and SCN.

This stage starts by training the best performing prediction models identified during the first evaluation stage (Fig. 3). The training is performed using the subset of the Stack Overflow dataset not used in the first stage, which accounts for about 1 M answers overall. Then, the models are tested on each of the four test sets. The same set of scalar measurements and graphical evaluations used in the first assessment stage are also applied at this stage to assess and compare the performance of the classifiers.

---

[15] https://cran.r-project.org/package=caret

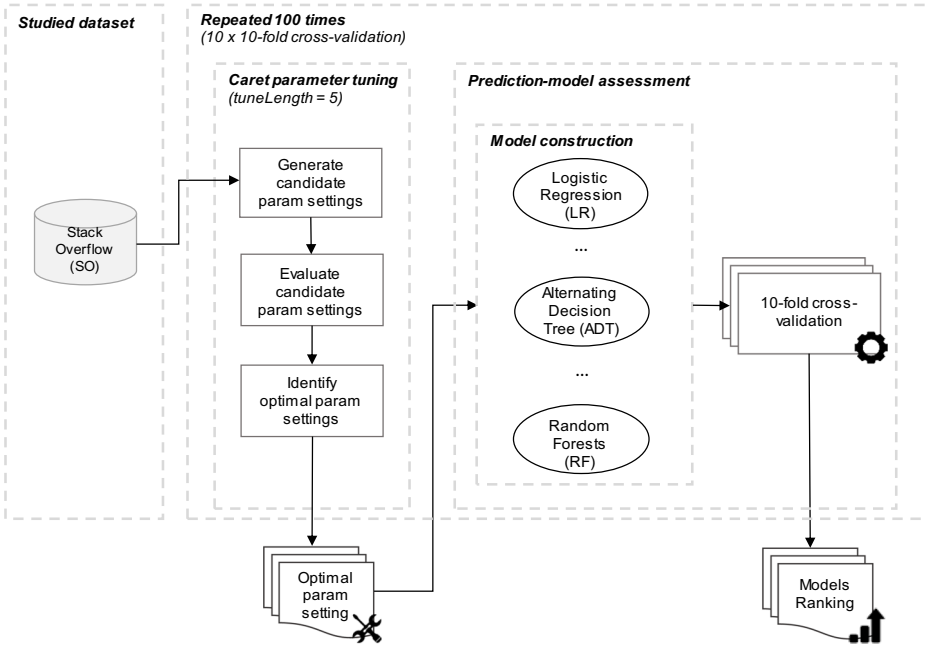

**Fig. 2** Workflow of the first assessment stage: best-answer prediction within Stack Overflow

## 6 Results

In this section, we present the results with respect to the two research questions.

### 6.1 Best-answer prediction within stack overflow

Table 8 reports the results of the Scott-Knott ESD test on the mean AUC values collected from the first evaluation stage, with 19 statistically distinct clusters of classifier performance. The 4 best prediction models in the top 3 clusters from the Scott-Knott ESD test, namely xgbTree, pcaNNet, earth, and gbm, achieve a mean AUC performance of about 0.9 and a maximum value ranging between 0.93 and 0.94, as also visible in Fig. 4, which shows the box plot of the AUC values collected during the repeated cross-validation process. On average, a large disparity in performance is observed between these four best models (mean AUC ≈ 0.9) and the worst (AUC = 0.59 for JRip), corresponding to a ~43 percentage points difference. All the prediction models proved to be stable in terms of AUC, given the small standard deviations values observed in this dataset (see Table 8).

Accordingly, in the following, we focus on these 4 classification techniques from the top 3 clusters. Table 9 reports their optimal parameter configurations with respect to maximizing AUC, and the overall amount of time taken by the parameter tuning process. The complete table is available online as additional material.[16]

First, as per the tuning runtime of the 4 models in the top 3 clusters, caret was able to complete the process in 2 to 8 h. The process took a few hours, often minutes, to complete also for the other classifiers, thus confirming that parameter tuning is computationally practical with respect to our experimental setting. There are a few notable exceptions though, such as knn, AdaBoost, and mlpWeightDecay, which instead took some days (85-125 h). Surprisingly, despite the long tuning process, none of these classifiers performed very well.

---

[16] https://github.com/collab-uniba/emse_best-answer-prediction/tree/master/additional_material

**Table 7** Overview of the 26 studied classifiers and their tuned parameters (the '?' indicates a yes/no parameter)

| Family | Classifier (short name) | Parameters | Description |
| --- | --- | --- | --- |
| Regression-based | Generalized Linear Models (glm) | – | |
| | Multivar. Adaptive Regression Splines (earth) | degree | Max degree of interaction |
| | | nprune | Max # of terms in model |
| Bayesian | Naïve Bayes (nb) | fL | Laplace correction factor |
| | | usekernel? | Use kernel density estimate |
| | | adjust | Bandwidth adjustment |
| Nearest Neighbor | K-Nearest Neighbor (knn) | k | # Clusters |
| Discrimination Analysis | Linear Discriminant Analysis (lda) | – | |
| | Penalized Discriminant Analysis (pda) | lambda | Shrinkage penalty coefficient |
| | Flexible Discriminant Analysis (fda) | degree | Max degree of interaction |
| | | nprune | Max # of terms in model |
| Decision Trees | C4.5-like trees (J48) | C | Confidence factor for pruning |
| | Logistic Model Trees (LMT) | iter | # Iterations |
| | Classification and Regression Trees (rpart) | cp | Complexity penalty factor |
| Support Vector Machines | SVM with Linear Kernel (svmLinear) | C | Cost penalty factor |
| Neural Networks | Standard (nnet) | size | # Hidden units |
| | | decay | Weight decay penalty factor |
| | Feature Extraction (pcaNNet) | size | # Hidden units |
| | | decay | Weight decay penalty factor |
| | Model Averaged (avNNet) | bag? | Apply bagging at each iteration |
| | | size | # Hidden units |
| | | decay | Weight decay penalty factor |
| | Multi-layer Perceptron (mlp) | size | # Hidden units |
| | Voted-MLP (mlpWeightDecay) | decay | Weight decay penalty factor |
| | | size | # Hidden units |
| | Penalized Multinomial Regression (multinom) | decay | |
| Rule-based | Repeated Incremental Pruning to Produce Error Reduction (JRip) | NumOpt | # Optimization iterations |
| Bagging | Random Forests (rf) | mtry | # Predictors sampled |
| | Bagged CART (treebag) | – | |
| Boosting | Gradient Boosting Machine (gbm) | n.trees | # Trees to fit |
| | | interact. Depth | Max depth of var. interactions |
| | | shrinkage | Param. applied to tree expansion |
| | | n.minobsinnod | Min # terminal nodes |
| | Adaptive Boosting (AdaBoost) | mfinal | # Boosting iterations |
| | | maxdepth | Max tree depth |
| | | coeflearn | Weight updating coefficient |
| | General. Additive Models Boost (gamboost) | mstop | # Initial boosting iterations |
| | | prune? | Pruning w/ stepwise feat. Selection |
| | Logistic Regression Boosting (LogitBoost) | nIter | # Boosting iterations |
| | eXtreme Gradient Boosting Tree (xgbTree) | nrounds | Max # iterations |
| | | maxdepth | Max tree depth |
| | | eta | Step-size shrinkage coefficient |
| | C5.0 (C50) | trials | # Boosting iterations |
| | | model | Decision trees or rule-based |
| | | winnow? | Apply predictor feature selection |

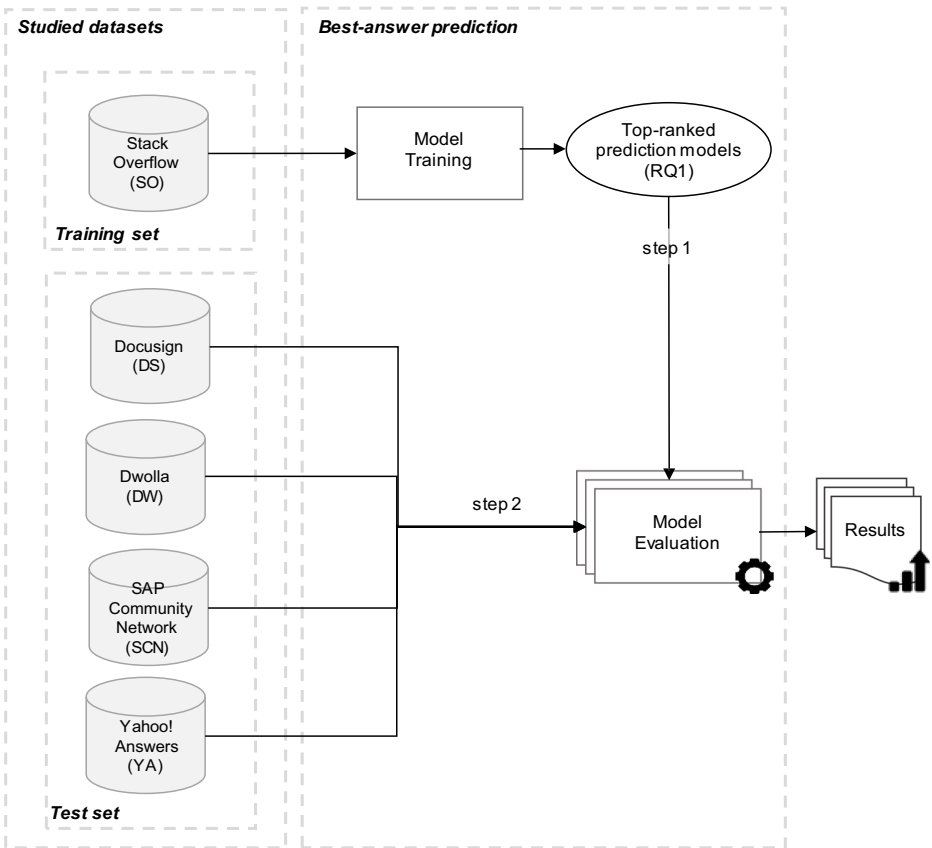

**Fig. 3** Workflow of the second assessment stage: cross-platform best-answer prediction

Second, we notice that for 21 out the 23 classifiers with parameters at least one of them has been modified from its default value in R during the parameter tuning step. This shows that the default R values for the studied classifiers are far from the optimal configuration in our case. Besides, performance with optimal parameter configuration improved in all four cases, with AUC percentage increase between 16 and 86%. In general, this happened for all the classifiers that have at least one parameter, with AUC percentage increase ranging between 2 and 113%. The only exception is LogitBoost, whose performance in both the tuned and the default configuration is the same. Overall, these results provide evidence that parameter tuning can largely improve the performance when building best-answer prediction models.

Finally, to identify further differences in the performance of the 4 models in the top 3 clusters, we performed a graphical assessment using ROC plot analysis. Yet, the ROC plot in Fig. 5 fails to reveal any differences. Instead, we observe that, while the differences between the four models are negligible in terms of AUC, the *Balance* metric shows slightly more marked performance disparities (see Table 10), with xgbTree and gbm coming out on top.

Accordingly, at the end of the first stage of evaluation, we answer RQ1 and select xgbTree and gbm as the two best-performing models for predicting best answers in Stack Overflow in terms of AUC and *Balance*.

**Table 8** Prediction models clustered and ranked according to the Scott-Knott ESD test. The top 3 clusters with AUC ≥ .90 are shown in gray

| Cluster | Prediction model | Mean AUC | Max | Min | SD |
| --- | --- | --- | --- | --- | --- |
| 1 | xgbTree | 0.91 | 0.94 | 0.88 | 0.02 |
| 2 | pcaNNet | 0.91 | 0.93 | 0.84 | 0.03 |
|   | earth | 0.91 | 0.93 | 0.83 | 0.03 |
| 3 | gbm | 0.90 | 0.94 | 0.83 | 0.04 |
| 4 | gabmboost | 0.83 | 0.88 | 0.79 | 0.03 |
| 5 | nnet | 0.82 | 0.83 | 0.80 | 0.01 |
| 6 | LMT | 0.82 | 0.84 | 0.79 | 0.02 |
|   | avNNet | 0.82 | 0.83 | 0.80 | 0.01 |
| 7 | C5.0 | 0.81 | 0.83 | 0.79 | 0.01 |
|   | AdaBoost | 0.81 | 0.82 | 0.79 | 0.01 |
| 8 | multinom | 0.81 | 0.84 | 0.73 | 0.03 |
|   | rf | 0.81 | 0.84 | 0.77 | 0.02 |
| 9 | lda | 0.80 | 0.84 | 0.74 | 0.03 |
|   | mlpWeightDecay | 0.80 | 0.84 | 0.71 | 0.04 |
| 10 | glm | 0.79 | 0.81 | 0.77 | 0.01 |
| 11 | nb | 0.79 | 0.81 | 0.76 | 0.02 |
|   | mlp | 0.79 | 0.80 | 0.77 | 0.01 |
| 12 | fda | 0.78 | 0.81 | 0.74 | 0.02 |
| 13 | pda | 0.78 | 0.80 | 0.74 | 0.02 |
| 14 | knn | 0.76 | 0.78 | 0.74 | 0.01 |
| 15 | LogiBoost | 0.74 | 0.76 | 0.71 | 0.01 |
| 16 | treebag | 0.73 | 0.79 | 0.65 | 0.05 |
| 17 | J48 | 0.71 | 0.73 | 0.67 | 0.02 |
|   | rpart | 0.71 | 0.73 | 0.70 | 0.01 |
| 18 | svmLinear | 0.64 | 0.67 | 0.62 | 0.02 |
| 19 | JRip | 0.59 | 0.67 | 0.57 | 0.04 |

**Feature selection** In the following, we analyze which of the 22 features contribute most to predicting best answers in Stack Overflow. Furthermore, through this analysis, we want to know whether feature ranking can boost prediction performance.

We perform feature selection using the R package Boruta[17] (Kursa and Rudnicki 2010), which uses a wrapper approach built around an ensemble method (i.e., Random Forest) where classification is performed by majority voting of multiple unbiased weak classifiers (i.e., decision trees). The importance measure of an attribute is assessed through the loss of accuracy in classification, using a given attribute and computed separately for all trees in the forest. Then, a Z score is computed by dividing the average loss by its standard deviation. Table 11 reports the results of the feature selection process. First, all the 22 features are retained. Second, we observe that the most important features belong to the set of *meta* features. In fact, *rating score*-ranked (pos. 1, Z = 134.83), *rating score* (pos. 2, Z = 118.24), *age*-ranked (pos. 3, Z = 74.96), and *age* (pos. 5, Z = 57.11) are all in the top 5 of the most important features. Such *meta* features measure, respectively, the answer appreciation and speed. We also note that the feature that contributes the least to best-answer prediction is by far *contains hyperlinks* (Z = .02, pos. 22), followed by the *vocabulary* feature $LL_n$, both ranked and non-ranked (pos. 21 and 20, respectively), despite being the most computationally-intensive one in our feature set.

---

[17] https://m2.icm.edu.pl/boruta

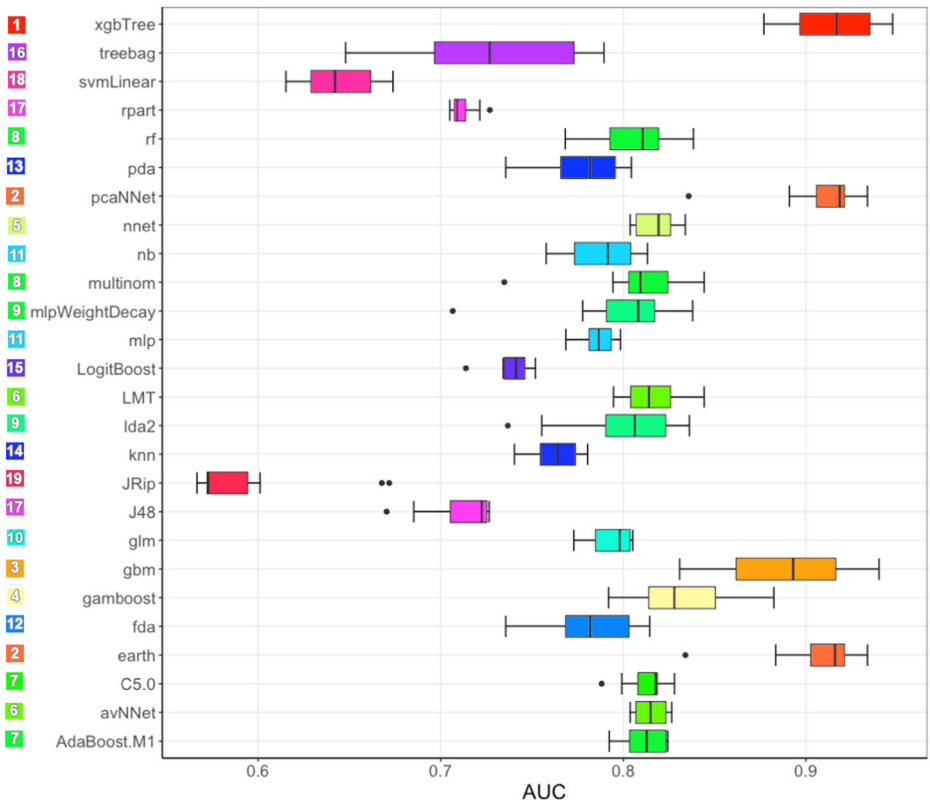

**Fig. 4** Box plots of classifiers' AUC performance values achieved during the parameter tuning stage on the Stack Overflow subset (numbers next to model names indicate their Scott-Knott ESD cluster)

Furthermore, we note that 8 features in the top 10 positions are ranked features, except for *rating score* and *age*. Finally, we underline the benefits of feature ranking for boosting prediction performance. Such benefits are evident when looking at the improvement due to ranking, measured as the relative change in positions or importance (Z) between the ranked and non-ranked version of a given feature (see the two rightmost columns in Table 11). For example, *Length* and *Length_ranked* are in position 16 and 9, respectively. The improvement due to ranking, in this case, is measured as a gain of +7 = − (9–16) positions; likewise, in terms of feature importance (Z) the improvement is +9.85 = 48.49–38.64, meaning that the ranked feature is deemed more important than the non-ranked counterpart by the feature selection algorithm. Overall, we note that all the ranked features in the table score better than their non-ranked counterparts, with an improvement in terms of positions ranging from +1 to +15 (+5 on average). Likewise, the improvement in terms of importance engendered by feature ranking varies between +1.22 and + 28.09 (+11.92 on average).

**Table 9** Default and optimal parameter configuration with AUC performance and overall tuning runtime (in hours) for the 4 models in the top 3 clusters

| Prediction Model | Default parameter configuration | Optimal parameter configuration | Overall tuning runtime | Default AUC performance | Optimal AUC performance |
|---|---|---|---|---|---|
| xgbTree | nrounds = 100<br>max_depth = 1<br>eta = 0.3 | nrounds = 200<br>max_depth = 4<br>eta = 0.1 | 6 h 47 m | .81 | .94 (+16%) |
| pcaNNet | size = 1<br>decay = 0 | size = 7<br>decay = 0.1 | 2 h 20 m | .50 | .93 (+86%) |
| earth | nprune = NULL<br>degree = 1 | nprune = 15<br>degree = 1 | 3 h 53 m | .50 | .93 (+86%) |
| gbm | n.trees = 100<br>interaction.depth = 1<br>shrinkage = 0.1<br>n.minobsinnode = 10 | n.trees = 250<br>interaction.depth = 3<br>shrinkage = 0.1<br>n.minobsinnode = 10 | 8 h 44 m | .79 | .94 (+19%) |

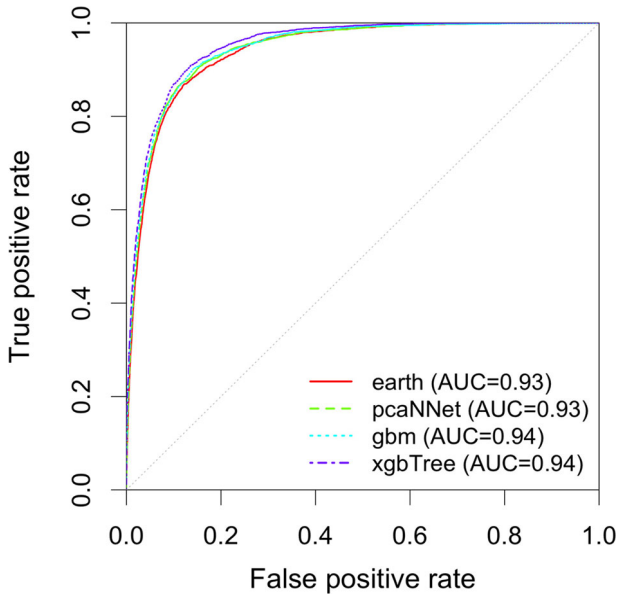

**Fig. 5** ROC plot for the 4 top-ranked classifiers

**Influence of score-related features** The Z scores from Table 11 indicate that score-related features are by far the strongest predictors in the models, so much so that one may argue that the models may just be selecting the answers with the highest score as the best answer. Therefore, to assess the extent to which the models depend just on upvotes, we re-build the classification models for the 4 top-ranked algorithms after excluding the rating score and rating score_ranked features. We use the dataset of ~341 K answers used in the tuning stage and optimal parameter configuration. For the sake of completeness, we note that we tested and obtained the same results using 10-fold cross-validation instead of performing a 70/30 training and test, re-using the tuned parameters.

**Table 10** Scalar measures of best-answer prediction performance within Stack Overflow for the classifiers in the top 3 clusters with optimal parameter configuration

| Models | AUC | Balance |
|---|---|---|
| xgbTree | 0.94 | 0.87 |
| pcaNNet | 0.93 | 0.85 |
| earth | 0.93 | 0.82 |
| gbm | 0.94 | 0.87 |

**Table 11** Result of feature selection on Stack Overflow (darker is better). The first position is assigned to the feature with the largest Z value, the last position to the smallest. The two rightmost columns show the extent to which each ranked feature improve over its non-ranked counterpart in terms of positions and Z

| Feature category | Feature name | Position | Feature importance (Z) | Improvement of ranked feature(as compared to non-ranked ver.) | |
|---|---|---|---|---|---|
| | | | | Positions | Importance (Z) |
| Linguistic | Length | 16 | 38.64 | +7 | +9.85 |
| | Length_ranked | 9 | 48.49 | | |
| | Word count | 14 | 40.06 | +3 | +2.92 |
| | Word count_ranked | 11 | 42.98 | | |
| | No. of sentences | 13 | 41.7 | +5 | +7.57 |
| | No. of sentences_ranked | 8 | 49.34 | | |
| | Longest sentence | 18 | 38.21 | +11 | +13.83 |
| | Longest sentence_ranked | 7 | 50.29 | | |
| | Avg. words per sent. | 12 | 42.27 | +6 | +12.58 |
| | Avg. words per sent._ranked | 6 | 54.85 | | |
| | Avg. chars per word | 19 | 36.87 | +15 | +28.09 |
| | Avg. chars per word_ranked | 4 | 64.26 | | |
| | Contains hyperlinks | 22 | 0.02 | N/A | N/A |
| Meta | Age | 5 | 57.11 | +2 | +17.85 |
| | Age_ranked | 3 | 74.96 | | |
| | Rating score | 2 | 118.24 | +1 | +16.59 |
| | Rating score_ranked | 1 | 134.83 | | |
| Vocabulary | $LL_n$ | 21 | 12.14 | +1 | +1.22 |
| | $LL_n$_ranked | 20 | 13.36 | | |
| | F-K | 15 | 39.55 | +5 | +8.72 |
| | F-K_ranked | 10 | 48.27 | | |
| Thread | Answer count | 17 | 38.63 | N/A | N/A |

Results are reported in Fig. 6, which shows that the AUC performance achieved without the two features ranges between .81 and .82. The comparison of these results against the performance of the full-fledged models developed during the first stage of evaluation (AUC = .93–.94, see Table 10), suggests that excluding the rating score and rating score_ranked features causes a ~13% decrease in prediction performance.

**Analysis of corner cases** To garner a deeper understanding of the behavior of the prediction models, we perform a quantitative analysis of corner cases occurring on particular values for the predicting variables. Specifically, we focus on studying the behavior of the 4 top-ranked models when answer upvote score is not a discriminant feature, that is: (i) when all the answers in a question thread have no upvotes; (ii) when all the answers in a question thread have the same vote; (iii) when there is only one answer in the question thread. For all these three corner cases, we repeat the same evaluation performed earlier.

In the first two cases, the datasets are created artificially by setting the rating score feature for all the answers in the ~341 K answers dataset, respectively, to 0 and the median value. The respective AUC performances of the four models (.81–.82) are either identical or have negligible differences (i.e., ~0.01) compared to those achieved when the rating score and rating score_ranked predictors are simply removed. This is because features with the same value, whether 0 or positive, exhibit null variance and, therefore, provide no information to learn from.

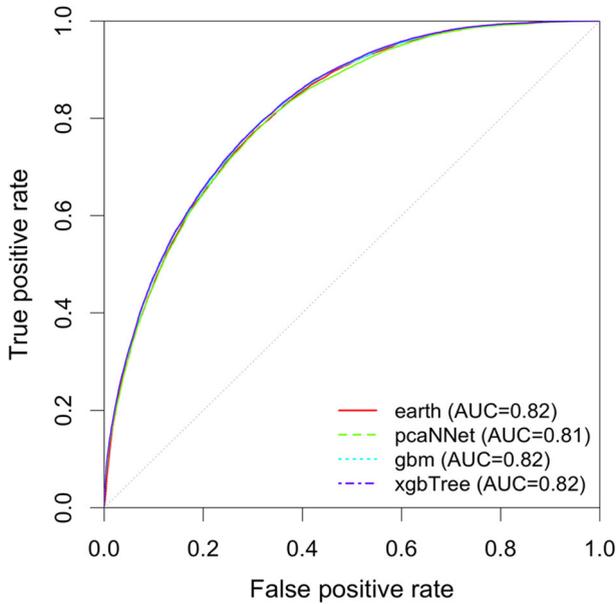

**Fig. 6** ROC plot for the 4 top-ranked classifiers without using the upvotes-related features

As for the third case, we filter the same ~341 K answers dataset, keeping only the question threads with just one answer received (n = ~8 K). The evaluation shows consistent prediction performances of AUC = ~0.70 for all the four models.

**Qualitative analysis** To further our understanding of *why* some classification algorithms perform better than others, we also conduct a qualitative analysis of the predictions on a random subset of 380 answers, corresponding to the ~95% interval for the whole set of 341 K answers used at this stage. Our analysis process is the following. We first identify the common cases of correct classifications and misclassifications between the 4 best performing models in the top 3 clusters. Then, we do the same for the worst performing classification algorithms in the bottom-3 clusters (i.e., JRip, svmLinear, rpart, and J48). Finally, to uncover patterns, we identify which answers classified correctly by the best models are misclassified by the worst models. Among the misclassified instances, we notice the recurring presence of cases where the accepted solution is not the fastest. For example, for question 20,864,634[18] "*String.Format Not Aligning Correctly using ToShortDateString and String,*" most of the models select the fastest answer[19] (i.e., false positive) instead of the actual accepted solution[20] (i.e., false negative). A similar scenario can be observed for question 27,727,589[21] "*__cplusplus compiler directive defined and not defined*." Albeit to a much lesser extent, we also observe the presence of misclassifications when the accepted answer is not the most upvoted. For example, for question 27,727,433[22] "*How is BareMetalOS allocating memory in Assembly without malloc, brk, or mmap?*", the poor performing models typically fail in recommending as candidate best answer the one with 4 upvotes[23] (i.e., false positive) while ignoring the actual accepted solution with just one upvote[24] (i.e., false negative). In other words, we notice that one reason accounting for the difference in performance between the classifiers is the extent to which the built models managed to learn from other features, without relying extensively on the features in the *meta* category, in particular, the *age* feature. We doublecheck this pattern by verifying that this tendency to misclassify such cases decreases with the other classifiers in the remaining 'middle' clusters. As for the other misclassified instances, there does not seem to be an underlying pattern to explain them.

---

[18] https://stackoverflow.com/questions/20864634
[19] https://stackoverflow.com/a/20864795
[20] https://stackoverflow.com/a/20864807
[21] https://stackoverflow.com/questions/27727589

**Timewise analysis** In our approach, best-answer prediction models learn what constitutes a good answer from random training and test sets that are stratified with respect to the proportion of accepted vs. non-accepted answers as in the original dataset. One possible issue with this approach, however, is that it disregards the intrinsic timeliness involved in the question answering process – that is, one may argue about including future answers in the training set for predicting the best answer to question threads in the test set, which actually happened earlier in the timeline.

In order to assess the influence of time ordering on best-answer prediction, we conduct a time-wise analysis, inspired by the batch forecasting approach for time series analysis (Hyndman and Athanasopoulos 2017). We use the same random subset of ~341 K answers used earlier, sorted by date. We found that content in the subset was contributed between July 2008 and January 2016. Then, we conduct a series of best-answer prediction experiments by splitting the subset into multiple, time-ordered training and test sets as follows.

At the first iteration (see Fig. 7a), all the posts from the entire subset $S$ that were contributed in 2008 are selected and included in the base training set $Train_0$. Then, we select from $S$ a first batch of unseen answers $Test_0$ that are not in $Train_0$ (i.e., $Test_0$ is selected from $S-Train_0$). We found the minimal time-window size for batch selection to be three weeks. Any batch smaller than that would lead to cases where AUC could not be computed due to the lack of answers predicted as solutions by the classification models. Therefore, $Test_0$ contains the answers contributed in the first three weeks of the set $S-Train_0$. Accordingly, the first prediction experiment involved training a model on $Train_0$ and testing it on $Test_0$. Performance was assessed using AUC.

When selecting the next batches, to generate more observations, we decided to shift the time window forward by two weeks, thus creating a one-week overlap between one batch and the next one (see Fig. 7b). Therefore, at the second iteration, the training set $Train_1$ contains $Train_0$ (i.e., the one from the first iteration) plus the content from week 1 and 2 of the previous test set (i.e., $Train_1 = Train_0 \cup \{week_1, week_2\}$) and the test set is $Test_1 = \{week_3, week_4, week_5\}$. To generalize, at the $i$-th iteration (see Fig. 7c), the training set is $Train_{i-1} = Train_0 \cup Train_1 \cup \ldots \cup \{week_{j-2}, week_{j-1}\}$, whereas the test set is $Test_{i-1} = \{week_j, week_{j+1}, week_{j+2}\}$.

We performed this timewise analysis using the two classifiers selected at the end of the first stage, i.e., xgbTree and gbm. In Fig. 8, we report the plot for xgbTree, whereas the other is available online as supplementary material in the GitHub repository.[25] The figure shows the plot of the AUC for each prediction experiments on the three-week batches of answers (i.e., the time-ordered test sets), plus the increasing size of the training set per experiment, normalized between 0 and 1. From the plot, we observe that, regardless of the training set size, the performance of the classifiers in terms of AUC is rather stable. The classifier performance appears to deviate from the average only when tested on the last batches. However, this is most likely because the most recent questions have had less time to receive answers. Finally, we also note that the average AUC for xgbTree is .91, which is the same value observed during the parameter tuning stage (see Table 8). Identical observations can be made for the gbm classifier which also achieves the same value of mean AUC = .91 in both conditions.

---

[22] https://stackoverflow.com/questions/27727433
[23] https://stackoverflow.com/a/27727523
[24] https://stackoverflow.com/a/27729675

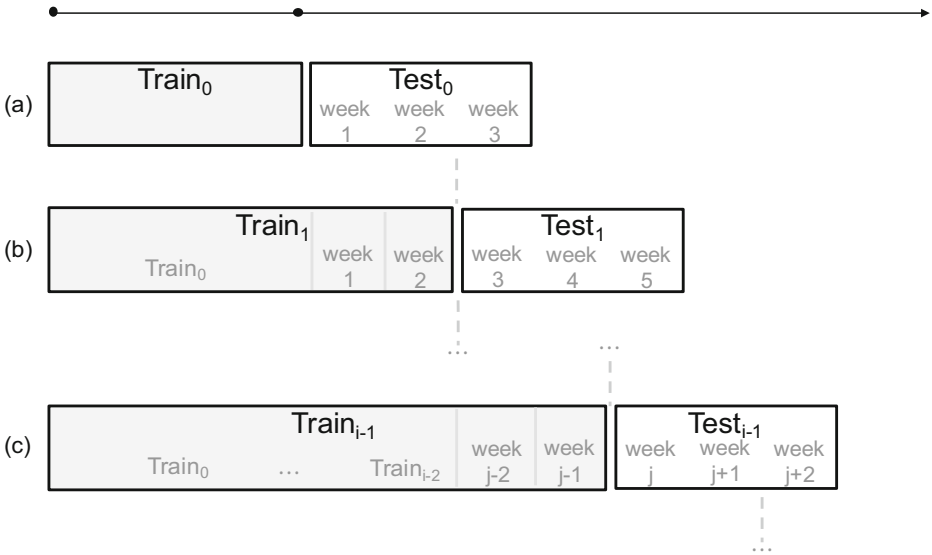

**Fig. 7** The process used for selecting multiple, time-ordered training and test sets, with a three-week time window for batch selection and a shift forward by two weeks at each iteration

Accordingly, we conclude that our approach to best answer prediction is sound regardless of the time ordering of the datasets.

### 6.2 Cross-platform best-answer prediction

In this section, we report the results from the cross-platform prediction stage of the study. Specifically, in Figs. 8a, b and c we report the ROC plots. For each of the four test sets, we use the method of DeLong et al. (1988) to measure the statistical significance of the difference between the areas under the two ROC curves. The test was performed using the pROC package available in R (Robin et al. 2011). We report the Z statistic, the $p$-value, and the power of test.

First, with respect to the legacy forum DocuSign, the ROC plot analysis (Fig. 9a) shows that the gbm curve clearly dominates xgbTree. Consistently, gbm outperforms xgbTree in terms of AUC values (.73 vs. .67) and *Balance* (.64 vs .61, see row I in Table 12). The ROC test confirms that the difference is statistically significant ($Z = 14.34$, $p < 0.01$). Regarding the Dwolla legacy forum, instead, no curve dominates the other in the ROC plot (Fig. 9b) although gbm performs slightly better than xgbTree in terms of *Balance* (.65 vs. .62) and AUC (.72 vs. .70, see row II in Table 12). However, the ROC test has not enough power (.20) to assess the difference between the two areas reliably, due to the limited size of the Dwolla test set.

---

[25] https://github.com/collab-uniba/emse_best-answer-prediction/tree/master/additional_material

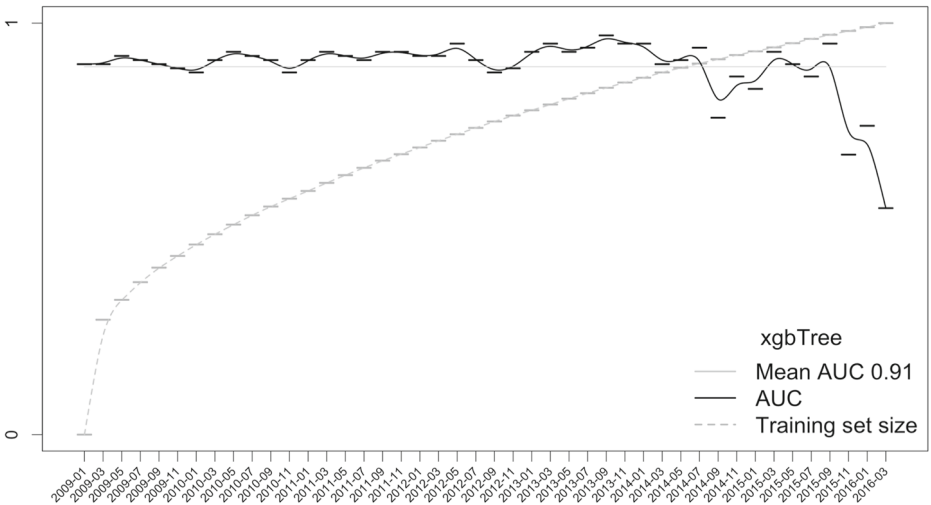

**Fig. 8** Plot of the xgbTree classifier performance (AUC) for the timewise analysis

With respect to modern platforms, we find that xgbTree outperforms gbm in the case of Yahoo! Answers, whereas the latter is better at predicting best answers from the SCN test set. More specifically, in the case of Yahoo! Answers, the ROC plot shows that the xgbTree curve dominates the other (Fig. 9c). In fact, xgbTree outperforms gbm with respect to AUC (.74 vs .71, see row III in Table 12). The difference between the two areas is also statistically significant ($Z = -30.86$, $p < 0.01$). Instead, the two prediction models have almost the same performance in terms of *Balance* (.64 vs .65, see row III in Table 12). Regarding SCN, we observe that the gbm model dominates xgbTree in the ROC space (Fig. 9d) and that the difference between the two AUCs (.71 vs .68) is statistically significant ($Z = 21.48$, $p < 0.01$). Consistently, gbm outweighs xgbTree in terms of *Balance* (.65 vs .62, see row IV in Table 12).

Furthermore, we observe that the values of the scalar metrics in Table 12 are close to the average with small standard deviation observed. This suggests that xgbTree and gbm models behave consistently across both modern and legacy Q&A platforms, regardless of the different pos/neg ratio of the test sets.

To assess the goodness of model performance in the cross-platform setting, we define a benchmark with two different reference models. Specifically, we create a *baseline* as the performance of a trivial rejector (i.e., a model always predicting the 'not accepted' majority class,) on the four datasets other than Stack Overflow in our study. We also define an *upper bound* as the performance of the xgbTree and gbm models trained and tested on each of these datasets (i.e., the within-platform prediction setting for each dataset), with automated parameter tuning and repeated 10-fold cross-validation. Results are reported in Table 13 in terms of *Balance* and AUC. We observe that the performances of the cross-prediction models 'sit in between' those of the two reference models.

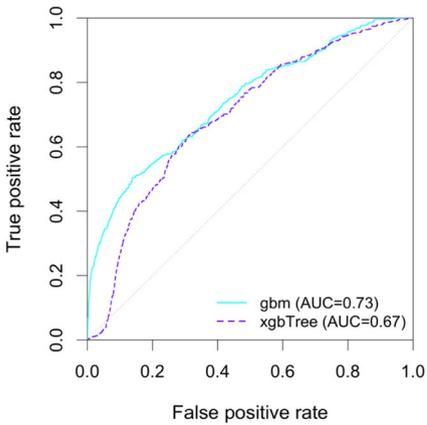
(a) DocuSign (Z=14.34, p<0.01, power=1)

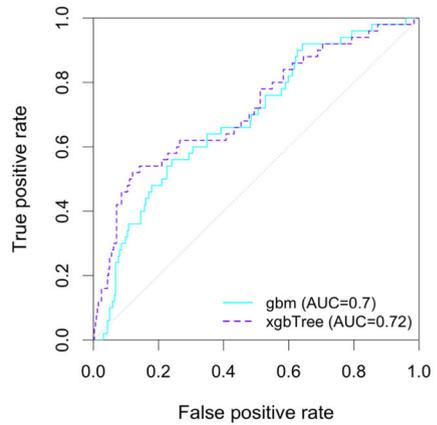
(b) Dwolla (Z=-2.65, p<0.01, power=.20)

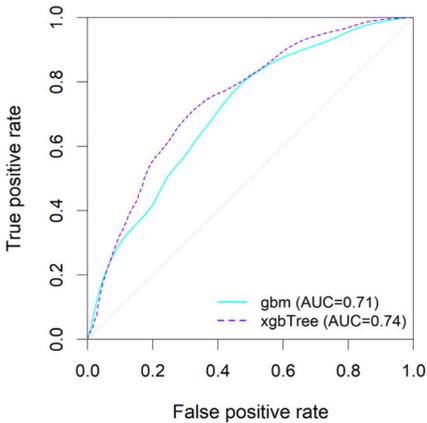
(c) Yahoo! Answers (Z=-30.86, p<0.01, power=1)

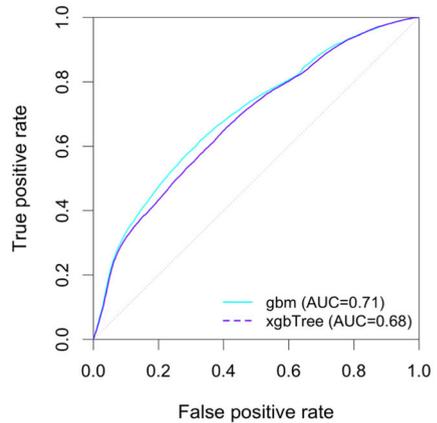
(d) SAP Community Network (Z= 21.48, p<0.01, power=1)

**Fig. 9** ROC plots for the two selected classification models trained on the Stack Overflow and tested on the four test sets. For each test set, the result of DeLong's ROC test statistic is reported, along with its significance level and power of test

Then, a two-tailed Wilcoxon signed-rank test is presented in Table 14 to assess whether the performance observed in the cross-platform setting is significantly different from the baseline and the upper-bound ones. We observe that both xgbTree and gbm models in the cross-platform prediction setting perform significantly better than the trivial rejector baseline in terms of AUC ($W = 16$ and $p = 0.021$ for xgbTree, $W = 16$ and $p = 0.017$ for gbm) and *Balance* ($W = 16$ and $p = 0.029$ for xgbTree, $W = 16$ and $p = 0.026$ for gbm), and with observable differences, i.e., large effect size with Cohen's $|\delta|$ values larger than 0.8 (Cohen 1988). Finally, we observe that, regardless of the metric, the test fails to reveal a significant difference between the cross-platform xgbTree/gbm models and the within-platform upper-bound models (see Table 14).

At the end of the second stage of evaluation, we answer RQ2 by noting that the performances of xgbTree and gbm (the two best prediction models from the previous stage) are similar across the four test sets in the cross-platform setting and also comparable to the within-platform upper bounds.

Table 12 Scalar measures of cross-platform best-answer prediction performance for the two models selected from the top rank

|  | Test set (pos/neg) | xgbTree | | gbm | |
| --- | --- | --- | --- | --- | --- |
|  |  | AUC | Balance | AUC | Balance |
| I | Docusign (1:10) | 0.67 | 0.61 | 0.73 | 0.64 |
| II | Dwolla (1:7) | 0.72 | 0.62 | 0.70 | 0.65 |
|  | *Avg.* | *0.69* | *0.61* | *0.71* | *0.64* |
| III | Yahoo (1:4) | 0.74 | 0.64 | 0.71 | 0.65 |
| IV | SCN (1:15) | 0.68 | 0.62 | 0.71 | 0.65 |
|  | *Avg.* | *0.71* | *0.63* | *0.71* | *0.65* |
|  | *Overall S.D.* | *0.03* | *0.01* | *0.01* | *0.00* |

## 7 Related Work

Most previous work in mining community Q&A sites has focused on: (i) assessing the quality of questions and answers (Ponzanelli et al. 2014; Xia et al. 2016; Roy et al. 2017); (ii) understanding how software developers interact with each other on Q&A sites (Treude et al. 2011); (iii) providing empirical evidence on how to write good questions and answers (Bosu et al. 2013; Calefato et al. 2018); the impact of sentiment on getting an answer accepted (Calefato et al. 2015); (iv) the role played by social cues on the perceived quality of an answer (Hart and Sarma 2014); (v) the topics discussed by developers (Bajaj et al. 2014; Barua et al. 2012); (vi) retrieving semantically linked questions (Xu et al. 2016a; Xu et al. 2016b); and (vii) summarizing answers (Xu et al. 2017).

Table 15 summarizes the prior work reviewed next, which is strictly related to best-answer prediction for technical help requests.

Adamic et al. (2008) report the first study on best-answer prediction using data from the *Programming & Design* category of Yahoo! Answers. Their prediction model, built also on user-related features, achieved a prediction *Accuracy* of 73% by running a 10-fold cross-validation with logistic regression. Analogously, Shah and Pomerantz (2010) ran a 10-fold cross-validation with logistic regression using a dataset from Yahoo! Answers that is comparable in size to ours, achieving an Accuracy of 85%. Unlike our work, however, their dataset also contains non-technical content that is not allowed in Stack Overflow and the feature set contains again user-related features that we did not include because computationally intensive. Neither Adamic et al. (2008) nor Shah and Pomerantz (2010) verified the assumption of logistic regression about no collinearity between features (Fenton and Neil 1999).

The setting of the study by Tian et al. (2013) is more directly comparable to ours because they trained a classifier on a dataset obtained only from Stack Overflow without relying on user-related features. Besides, their feature set is similar to our set of non-ranked features. To evaluate their classification model, however, they ran a 2-fold cross-validation with Random Forest and achieved a prediction *Accuracy* of 72%.

**Table 13** Cross-platform model performance benchmarking (darker is better)

| Models | xgbTree | | | | | gbm | | | |
|---|---|---|---|---|---|---|---|---|---|
| Datasets (pos/neg) | | Trivial rejector | Cross-platform model | Within-platform model | Cross- vs. within-platform models performance variation | | Trivial rejector | Cross-platform model | Within-platform model | Cross- vs. within-platform models performance variation |
| Docusign (1:10) | Balance | 0.36 | 0.61 | 0.37 | +65% | | 0.36 | 0.64 | 0.39 | +64% |
| | AUC | 0.49 | 0.67 | 0.74 | −9% | | 0.49 | 0.73 | 0.75 | −3% |
| Dwolla (1:7) | Balance | 0.38 | 0.62 | 0.48 | +29% | | 0.38 | 0.65 | 0.43 | +51% |
| | AUC | 0.49 | 0.71 | 0.83 | −14% | | 0.49 | 0.71 | 0.83 | −14% |
| Yahoo (1:4) | Balance | 0.46 | 0.64 | 0.90 | −29% | | 0.46 | 0.65 | 0.96 | −32% |
| | AUC | 0.50 | 0.74 | 0.97 | −24% | | 0.50 | 0.71 | 0.96 | −26% |
| SCN (1:15) | Balance | 0.34 | 0.62 | 0.30 | +107% | | 0.34 | 0.65 | 0.77 | −16% |
| | AUC | 0.50 | 0.68 | 0.78 | −13% | | 0.50 | 0.71 | 0.77 | −8% |

Table 14 Results of the Wilcoxon signed-rank test of the performance measures (W statistics larger than critical value 10 are shown in bold)

| Measures | xgbTree cross-platform model vs. | | | | gbm cross-platform model vs. | | | |
| --- | --- | --- | --- | --- | --- | --- | --- | --- |
| | Trivial rejector W | $p$-value (Cohen's δ) | Within- platf. W | $p$-value (Cohen's δ) | Trivial rejector W | $p$-value (Cohen's δ) | Within- platf. W | $p$-value (Cohen's δ) |
| AUC | 16 | 0.021* (6.32) | 0 | 0.100 (−1.86) | 16 | 0.017* (21.50) | 0 | 0.100 (−1.12) |
| Balance | 16 | 0.029* (5.66) | 7 | 0.583 (0.43) | 16 | 0.026* (5.12) | 8 | 0.361 (0.55) |

*denotes significance at 5% level

None of the above studies reported the dataset pos/neg ratio nor considered the effect of class imbalance on the classification performance, as these studies report the classification performance using exclusively *Accuracy*, a metric that is considered unreliable for dealing with imbalanced datasets.

The issue of using unreliable metrics is also found in other studies. Cai and Chakravarthy (2011) included a resampled dataset from Stack Overflow. For each question thread, they extracted five answers, the best answer plus four non-accepted answers, randomly selected. Although not specifically reported, we can deduce that the pos/neg ratio of their dataset is 1:4. Their classification model, built using 10-fold cross-validation with SVM, showed a low performance in terms of *Precision* (.55). Again, as in the case of *Accuracy*, this result cannot be considered reliable because *Precision* alone is not an adequate measure in presence of class imbalance.

Burel et al. (2012) compared the performance of their classifier in two different Q&A platforms, namely SAP Network Community and Stack Exchange, from which they picked Server Fault as the specific technical site. They found ADT to be the best performing classifier after a comparison against other decision trees-based learning algorithms (i.e., J48, Random Forest, Alternating Tree, and Random Trees). Rather than relying on a single scalar measure, they reported their findings using several measures, specifically *Precision* (.83–.85), *Recall* (.84–.85), *F-measure* (.83–.84), and AUC (.89–.92). However, the authors neither reported the pos/neg ratio of the dataset nor investigated the effect of class imbalance on the classification performance. Furthermore, they did not make use of any plot for the visual analysis of performance. Although the comparison of our results is difficult because they do not employ separate training and test sets (only within-site cross-validation) and rely on user-related features that we excluded, we achieved similar performance when predicting best answers within Stack Overflow (AUC up to .94). Furthermore, both studies provide consistent evidence on the poor performance of the *vocabulary* features used to assess answer readability.

Shah (2015) built a Bayesian Network model for best-answer prediction on Yahoo! Answers, using a limited set of textual and user-related features (12 in total). The textual features extracted from answers are the Flesch-Kincaid readability score, an entropy-based clarity score, the number of misspelled words, and the length in characters, words, and sentences; from questions, they extract the number of questions marks and the number of distinct words over the total number of words used in a question. Findings were reported using both single and aggregate scalar metrics (*Accuracy* = 89.2, *Precision* = .97, *Recall* = .86, AUC = .98). This work also includes a ROC plot for visualizing the classification performance. Compared to the other studies, the high performance of Shah's classifier can be arguably

**Table 15** Breakdown of related work on best-answer prediction for technical help requests

| Reference | Dataset (# question/answers) | pos/neg ratio | Feature categories (total #) | Feature ranking? | Experimental setting | Param tuning? | Other classifiers compared | Performance results | Graphical assessment |
|---|---|---|---|---|---|---|---|---|---|
| Adamic et al. (2008) | Yahoo! Answers – Programming & Design (N/A) | N/A | user, thread, linguistic (Anderson et al. 2012) | No | 10-fold cross-validation with Log. Regression | No | No | Acc =~73% | No |
| Shah and Pomerantz (2010) | Yahoo! Answers** (~1.3 K/5 K) | N/A | user, thread, meta, linguistic (D'Ambros et al. 2012) | No | 10-fold cross-validation with Log. Regression | No | No | Acc =~84% | No |
| Cai and Chakravarthy (2011) | Stack Overflow (1 K/5 K)* | 1:4 | textual, user (Davis and Goadrich 2006) | No | 10-fold cross-validation with SVM | No | No | P = .55 | No |
| Tian et al. (2013) | Stack Overflow (~103 K/196 K) | N/A | thread, meta, linguistic (Calefato et al. 2016) | No | 2-fold cross-validation with Random Forests | No | No | Acc =~72% | No |
| Burel et al. (2012) | SCN†(~95 K/427 K) Server Fault (SF)‡ (~36 K/95 K) | N/A | user, thread, meta, linguistic, vocabulary (19†/23‡) | No | 10-fold cross-validation with ADT | No | J48, Random Forest, Random Trees | P = .83, R = .84, F = .83 AUC = .88 (SCN) P = .85, R = .85, F = .80 AUC = .91 (SF) | No |
| Shah (2015) | Yahoo! Answers** (23 K Q/A pairs)* | 1:4 | textual, user (Bosu et al. 2013) | No | 70/30% training/test set split with Bayesian Network | No | No | Acc = 89.2, P = .97, R = .86, AUC = .98 | ROC plot |
| Gkotsis et al. (2014, 2015) | 21 Stack Exchange sites (incl. Stack Overflow)** (~3 M/7 M) | N/A | thread, meta, linguistic, vocabulary (Cai and Chakravarthy 2011) | Yes | 10-fold cross-validation with ADT Cross-site leave-one-out with ADT | No | Yes (unspecified) | P = .82, R = .66, F = .73 AUC = .85 (SO only) P = .84, R = .70, F = .76 AUC = .87 (avg) | ROC plot |
| Calefato et al. (2016) | Stack Overflow (~91 K/~232 K) DocuSign (DS) (~1.5 K/~4.7 K) | | thread, meta, linguistic, vocabulary (Davis and Goadrich 2006) | Yes | Cross-site training vs. test set with ADT | No | J48, Random Forest, Random Trees | Acc 89.6, P = .85, R = .90, F = .86, AUC = .71 (DS) | No |
| Zheng and Li (2017) | Stack Overflow (~706 K/2.1 M) | N/A | textual, user, linguistic, NLP, IR (Fu et al. 2016) | No | Unspecified with custom ensemble classifier | No | No | P = .63, R = .59, F = .61 | No |
| This study | Stack Overflow (SO) (507 K/1.37 M) DocuSign (DS) (~1.5 K/~4.7 K) | ~1:4 ~1:10 | thread, meta, linguistic, vocabulary (Davis and Goadrich 2006) | Yes | 10-fold cross-validation with 26 classifiers Cross-site training vs. test set | Yes | Yes (26 classifiers) | AUC = .94 (SO) AUC = .73, F = .82, G = .65, Bal = .64 (DS) | ROC plots |

**Table 15** (continued)

| Reference | Dataset (# question/answers) | pos/neg ratio | Feature categories (total #) | Feature ranking? | Experimental setting | Param tuning? | Other classifiers compared | Performance results | Graphical assessment |
|---|---|---|---|---|---|---|---|---|---|
| | Dwolla (DW) (103/375) | ~1:7 | | | | | | AUC = .71, F = .80, G = .65, Bal = .65 (DW) | |
| | Yahoo! Answers – Progr. & Design (YA) (~41.2 K/~105 K) | ~1:4 | | | | | | AUC = .74, F = .66, G = .65, Bal = .64 (YA) | |
| | SCN (~35.5 K/~141.7 K) | ~1:15 | | | | | | AUC = .71, F = .80, G = .65, Bal = .65 (SCN) | |

\* Opportunistically sampled for selecting question threads with 1 best answer and 4 non-accepted answers

\*\* Dataset mixes technical and non-technical help requests

explained by the quite different study setting resulting in a less challenging classification task. As in the case of Cai and Chakravarthy (2011), rather than the entire question threads, a small dataset of 23,000 question/answer pairs is built, with five possible answers for each question (i.e., with an artificial pos/neg ratio of 1:4).

Analogously to our study, Gkotsis et al. (2014, 2015) did not rely on user-related features while also making use of feature ranking. The authors employed a very large dataset consisting of 21 Stack Exchanges sites, including Stack Overflow, thus mixing both technical and non-technical help requests. The sets of features employed, mostly computationally cheap, are consistent with ours. Furthermore, in one of their study, they performed a leave-one-out validation on the remaining 20 sites, obtaining an average performance of AUC = .87.

Besides the different model-validation technique employed, the higher prediction score in their experiment can be explained by using only sister sites belonging to the Stack Exchange network of Q&A sites; in our case, instead, our cross-context experiment involved sites from completely distinct Q&A platforms, both modern and legacy.

Zheng and Li (2017) recently performed a study on Stack Overflow with a couple of distinctive traits compared to previous work: (i) the development of a custom ensemble classifier that stacks multiple AdaBoost base learners; (ii) the extraction of features from code snippets using measures typical of Information Retrieval (i.e., *tf-idf*) and Natural Language Processing (i.e., LDA). However, the performance of their classification model, expressed in terms of *Precision* (.63), *Recall* (.59), and *F-measure* (.61), is one of the lowest reported in the literature despite the use of computationally-intensive features.

Finally, the current study extends our previous work (Calefato et al. 2016) where we trained an ADT classifier on Stack Overflow to cross-predict best answers in DocuSign. We found that the model trained without the *meta* features (i.e., *age* and *rating score*), either ranked and non-ranked, achieves almost identical prediction performance (AUC≈.70) compared to when they are available. Because the set of features employed in our previous work is the same as in the current study, this finding suggests that our approach is robust and capable of coping with the lack of such pieces of information. Besides, we compared the performance of the ADT model against three naïve rule-based classifiers, which would select, respectively, (i) the most upvoted answer (i.e., with the highest *rating score*), (ii) the first answer received (i.e., with the smallest *age*), and (iii) either of the previous two. The three rule-based classifiers achieved good performance, accurately predicting the accepted solution in about 90% of the cases. However, their performance in the case of DocuSign was very poor, slightly above random prediction. Due to our intention to study cross-platform prediction more thoroughly, compared to (Calefato et al. 2016), here the number of datasets has more than doubled, increasing from 2 to 5. In fact, we employ 4 test sets retrieved from different Q&A platforms, both modern (i.e., Yahoo! Answers and SCN) and legacy (i.e., Dwolla and, again, DocuSign). Besides, the size of the Stack Overflow training set has increased from ~230 K to almost 1.4 M answers, a much more representative subset of the 23 M answers available as of this writing. Furthermore, in the current study, we assess 26 different classifiers, from different families, whereas in (Calefato et al. 2016) we dealt with only 4 classifiers, all belonging to the family of decisions trees. Lastly, in this study, we perform a graphical assessment of prediction performance and automated parameter tuning of the classifiers, which were both missing in (Calefato et al. 2016). In addition, we also replaced *Accuracy* with *Balance*, as a more reliable scalar metric of performance, while also relying on ROC plots for graphical assessment. Because of all these changes, the current paper provides the readers with a full framework for the reliable assessment of best-answer prediction performance in both within and cross-platform settings.

## 8 Discussion

In our study, we employed five different datasets and compared the performance of 26 classifiers using both aggregate scalar metrics and plot analysis, thus providing a reliable assessment of performance in presence of imbalanced datasets. Our evaluation focused on (i) using automated parameter tuning to identify the best-performing models with respect to Stack Overflow, and (ii) assessing them in a cross-context setting, where models trained on Stack Overflow are tested on other technical Q&A platforms, both modern and legacy.

**Best-answer Prediction within Stack Overflow** Regarding the first research question, we confirm the findings from previous research on defect prediction (Ghotra et al. 2015; Hall et al. 2012; Jiang et al. 2008b; Mende and Koschke 2009; Tosun and Bener 2009) that the classifier choice and parameter tuning both have a large impact on prediction performance. In fact, assessing the performance of the 26 prediction models within Stack Overflow, we found 19 statistically distinct clusters (see Table 8), and observed more than a 40% increase in average AUC performance between the best (xgbTree) and the worst ranked (JRip) classification techniques, according to the Scott-Knott ESD test.

We also confirm evidence provided by previous research (Fu et al. 2016; Tantithamthavorn et al. 2016) about automated parameter tuning being computationally practical, with the tuning process taking a few hours, if not minutes, for most of the classifiers, with optimal AUC performance increasing up to 113% from default parameter configuration in R. Albeit here we provide the optimal parameter configuration for each of the 26 classifiers in our study, we remind that automated parameter tuning needs to be repeated whenever the data source or goals are changed (Fu et al. 2016), for example, after moving from the context of this study to the analysis of a dataset of software bugs for defect prediction.

Regarding features, the wrapper-based approach performed showed the importance of upvotes on prediction performance, with rating_score and rating_score_ranked being the two most discriminating features (see the Z statistic score in Table 11). Through the qualitative analysis, we found that one reason accounting for the difference in performance of the prediction models is their ability to learn from all the features without relying extensively on the *meta* features, in particular age and, to a much lesser extent, rating score. This finding is also consistent with the results of the quantitative analysis on the influence of rating scores on classification. We found that the best models are not entirely dependent on upvotes only and, therefore, the contribution of the other features is not superfluous for their building. In fact, the performances of the models re-built using the 4 top-learning algorithms and an ad hoc dataset that does not include the two features (see Fig. 6) show a ~13% loss in terms of AUC compared to the full-fledged models (see Fig. 5, AUC = .93–.94). Currently, in Stack Overflow, there are some 2 M resolved questions out of 8 M (25%) whose accepted solution is not the top voted answer. In such cases, the models are particularly useful in that they can recommend potential, 'reasonable' solutions beyond just the upvote score as when browsing Stack Overflow. Besides, our analysis also showed that the built models achieve a good AUC performance (.70–.82) in corner cases where answer rating score is absent or not discriminant, such as questions threads where there is just one answer or multiple answers with no upvotes.

Furthermore, our findings also show that the process of feature ranking contributes to boosting the performance of the prediction models in Stack Overflow. In fact, as suggested by Laradji et al. (2015) and Menzies (2016), we performed wrapper-based feature selection to test what factors have the largest impact on predicting best answers (Table 11). The results show that the ranked features contribute to the correct classification of the answers in Stack Overflow more than those non-ranked. Indeed, 8 of the features in the top 10 are ranked. Comparing the importance (Z) of each ranked feature to its non-ranked counterpart, an improvement can be observed in all the possible cases. Feature ranking preprocessing increased the performance of the prediction models by reducing the variability in the distribution of the feature values. This finding is consistent with prior work on cross-project defect prediction, which found model performance to improve when selecting projects with predictor values with similar distributions (Turhan et al. 2009) or by log-transforming predictors (Zhang et al. 2016). As regards best-answer prediction, a related finding is reported by Gkotsis et al. (2014, 2015), who also employed a similar set of ranked features in a cross-context study that involved only sister sites belonging to Stack Exchange. Accordingly, their results show that their approach is generalizable within Stack Exchange. Instead, because our study involves sites from completely distinct Q&A sites, both modern and legacy, our results go further, suggesting that our approach may be also generalizable across all Q&A platforms.

Through a timewise analysis, we also showed that our approach ensures good classification performance regardless of whether the answers in the dataset are time-ordered. We argue that we can ignore the ordering of answers because we rely on features that are able to capture aspects of answer goodness (e.g., the number of sentences, the length in characters) while not being influenced by time. Even the *age* feature is not affected by the time ordering since it calculates the relative time interval between the posting of one answer and the question. Furthermore, all our features are also computationally cheap, except for $LL_n$, the most intensive one and, incidentally, the second least important predictor in our feature set with almost negligible predictive power (see Table 11). Therefore, given the performance achieved by our best classifiers (max AUC = .94 for xgbTree and gbm), we argue that there is no need to employ computationally expensive features such as *tf-idf* (Manning et al. 2008), which in this scenario would capture how important a word is to an answer in a whole dataset. Given the sheer amount of questions (~15 M) and answers (~23 M) already posted on a Stack Overflow today, the time necessary to compute the metric for each word in an answer is far from negligible. On top of that, the measure would need to be updated every time the corpus is substantially changed. With Stack Overflow, this happens on a daily basis – e.g., in the year 2016 about 9 K new answers were contributed per day.

Overall, regarding our approach to solving the best-answer prediction problem, we focused on tackling the problem of question starvation by seeking to recommend potential, 'reasonable' solutions for unresolved questions that have received answers. Other researchers have tested different approached to solve the same problem. In particular, Nie et al. (2017) have developed an approach to 'resolve' questions by recommending a list of potential solutions chosen among those answers provided in response to other similar, possibly duplicate questions. As such, unlike ours, their approach has the potential to 'resolve' also questions with no answers (e.g., new answers). However, we argue that their solution, which was tested in two non-technical Q&A sites, is hardly applicable to domains like Stack Overflow because: (i) technical questions tend to be very specific, therefore the best answer (i.e., solution) to one question would hardly work for another, unless the two were duplicates; besides, (ii) posting duplicate questions in Q&A site is considered a bad practice that is strongly discouraged; in fact, by closing down duplicates, Stack Overflow community members are rewarded for keeping the knowledge base tidy.

Finally, one may argue that our approach is naïve or simplistic because it does not consider any semantic relationship between the question and the answers – i.e., none of the features aims at verifying that the 'best answer' is talking about the problem/issue presented in the question. Ensuring topic alignment would require using techniques such as LDA (Blei et al. 2003), which would not scale up well with a very large and ever-growing dataset like Stack Overflow because it requires a preliminary, computationally-intensive step for training and optimizing the topic model. On the contrary, our approach to suggesting candidate best answers, possibly even more than one for the same question thread, achieves satisfactory performance in coping with the endemic problem of the lack of accepted answers that plagues Stack Overflow. Our analysis also showed that the built models achieve a good AUC performance (.70–.82) in corner cases where answer rating score is absent or not discriminant, such as questions threads where there is just one answer or multiple answers with no upvotes.

**Cross-platform Best-answer Prediction** Regarding the second research question, results show that the two best prediction models from stage one (i.e., xgbTree and gbm) perform similarly across the four test sets, confuting the expectation that the larger the pos/neg ratio measured for an imbalanced dataset, the harder the classification task. Our finding provides further supporting evidence for the work by Lopez et al. (2013) who ran a series of classification studies on several datasets order by growing pos/neg ratio. As in our case, they also observed no pattern of behavior for any range of class distribution, with poor and good results obtained in the case of both low and high pos/neg ratios.

Hosseini et al. (2017) have performed a meta-analysis on cross-project defect prediction, proving practical guidelines for researchers interested in assessing prediction model performance in cross- vs. within-project settings. The methodology adopted in this paper ticks all the applicable boxes, such as parameter tuning, feature selection, using statistical tests with effect size. Our analysis revealed that cross-platform best-answer prediction is a much harder task than within-site prediction. In fact, while the two best-performing classifiers achieved a performance of max AUC = .94 within Stack Overflow, their performances dropped during cross-platform prediction by 20 to 27 points, respectively, in Yahoo (AUC = .74) and DocuSign (AUC = .67) (see Tables 10 and 12). This result is not entirely unexpected as cross-context prediction has also been problematic for defect prediction models when trained and tested on distinct software projects, especially when belonging to different companies (Nam and Kim 2015; Peters et al. 2013; Turhan et al. 2009; Hosseini et al. 2017). However, the reasons for this performance drop are not entirely clear yet. We hypothesize that our assumption of domain similarity (i.e., Q&A about software development in a broad sense) may have been overly optimistic. With almost 1.4 M answers to analyze, the models trained on Stack Overflow may have ended up learning data regularities that were not (entirely) relevant in the cross-platform prediction stage. Thus, the size and the generality of the Stack Overflow training set may have increased the 'domain distance' from the test sets (Turhan et al. 2011). Future replications of this study will seek to reduce this distance between the training and the test datasets by leveraging topic modeling and related tags to select a more relevant and 'closer' subset of questions threads from the entire Stack Overflow dataset. Furthermore, future replication will seek to improve the performance by leveraging more computationally intensive semantic features (e.g., Jaccard index) to leverage the textual coherence between the content of a question and its answers (Scalabrino et al. 2016).

Furthermore, to understand the extent to which the performance drop is related to the limitations of the prediction models rather than intrinsic difficulty of learning from the four data sets, we benchmarked each of the cross-platform models against a baseline and an upper bound, respectively a trivial rejector and a within-platform model that was trained and tested on the same dataset. We found that the models built in the cross-platform setting outperform the trivial rejector baseline in terms of AUC and *Balance*. Besides, regarding the comparison with the upper bound, we note that no statistically significant difference is observed between the performance of the cross- and within-platform models, regardless of the platform type (see Table 13 and Table 14). This finding is ultimately important because it shows that our approach can build cross-platform models whose performance is close to the upper bound of within-platform models in case of both legacy and modern sites.

Finally, given the highly predictive capability of metadata-related features (i.e., rating score and age, see Table 11 on feature selection), one might argue that simpler approaches that do not require many features and machine learning might be employed to retrieve best answers – that is, why feed them into a predicting model once it is already known which answer is the fastest and which has the highest score in a question thread? In our previous work (Calefato et al. 2016), we reported a series of prediction experiments performed by selecting different sets of features at a time, thus showing that an ADT prediction model trained without *meta* features achieved almost identical performance to when those were all enabled. Furthermore, in (Calefato et al. 2016), we benchmarked the same 'complex' prediction model trained on Stack Overflow with all the features enabled, against a few rule-based, naïve classifiers that would select the best answer as (i) the top-voted answer, (ii) the fastest answer, and (iii) either of the two. When tested on Stack Overflow, these 'strawman' models successfully identified the accepted answer in up to ~90% of the cases, showing that *meta* features are highly reliable predictors in Q&A sites where rapid answers and the practice of upvoting are widely used and part of the community culture (Mamykina et al. 2011). However, when tested on legacy Q&A sites such as DocuSign, we found the naïve models to achieve a poor performance, with AUC values close to random prediction (between .54 and .59). Consistently, we argued that such poor results are explained by the low number (18%) of question threads in the DocuSign dataset containing at least one answer that has received upvotes. In the current study, however, the performance of the two complex models tested on the DocuSign dataset in the cross-platform setting is similar if not better than on the other test sets (i.e., Dwolla, Yahoo! Answers, and SCN, see Fig. 8), despite the larger numbers of question threads containing at least one upvoted answer. In addition, we note that the distribution of answers that have received at least one upvote in the 4 test sets are in fact are quite different than Stack Overflow (74%), ranging between 4% (DocuSign) and 30% (Dwolla). Hence, the combined results from our studies suggest that while complex classifiers based on machine learning may be arguably superfluous for Stack Overflow alone, best-answer prediction within and across other Q&A platforms is a more challenging task that does require the development of more complex predictive models with multiple features.

## 9 Threats to Validity

Regarding threats to internal validity, we acknowledge that the increase of the threshold limiting the space for automated parameter tuning may yield different results. However, previous research has concluded that even the exploration of a small parameter space can lead to a large increase in classification performance (Fu et al. 2016; Tantithamthavorn et al. 2016). Besides, we had to manually annotate the Dwolla test set, because of the lack of accepted solutions. We acknowledge that only the original asker would be able to select the actual accepted solution. We mitigated this threat by following a careful process of manual annotation, which was performed by two of the authors who inspected each question thread and marked as an accepted solution only the message acknowledged with explicit positive feedback by the asker. Furthermore, we acknowledge that the use of a wrapper-based technique (e.g., Random Forests variable importance) for feature selection represents a threat to validity since there are many different techniques, such as Correlation Feature Selection (Hall 1999) and ANOVA (Saeys et al. 2007). Yet, given the size of the datasets used in the study and the number of different classifiers compared, we reserve to evaluate feature selection techniques in the domain of best-answer prediction in our future work. Finally, we are aware that recent work on defect prediction has shown how the choice of model validation technique (i.e., repeated cross-validation, in this case) may impact the performance estimate (Tantithamthavorn et al. 2017). Given that the datasets used in the study are made publicly available, this limitation might be addressed in future independent replications.

The problem of class imbalance can affect both internal and construct validity. On one hand, because the minority class contains less-common cases holding the highest interest from a learning point of view, algorithms may generate suboptimal classification models when trained on imbalanced datasets (He and Garcia 2009; Japkowicz and Stephen 2002). On the other hand, instead, some scalar metrics, such as *Precision* and *Accuracy,* have been found to be inadequate due to their sensitiveness to class distribution (He and Garcia 2009; Zhang and Zhang 2007). In the study, we contrasted both effects by combining scalar and graphic metrics to compare the performance of multiple classifiers. Specifically, we addressed the threat to internal validity by including ensemble learners in the set of the 26 classifiers assessed in the study. Ensembles of classifiers are considered an effective solution to the class imbalance problem (Lopez et al. 2013). Alternative approaches to deal with the problem are resampling and cost-sensitive learning (Malhotra and Khanna 2017). *Resampling* techniques generate a new balanced dataset by either eliminating random instances from the majority class (undersampling*)* or creating new 'synthetic' ones in the minority class (oversampling). Previous research has observed that balancing a dataset, albeit a practical approach, may alter the underlying statistical problem (Turhan 2012) and does not always translate into classification-performance gain (Lopez et al. 2013). *Cost-sensitive learning* approaches, instead, try to minimize the overall error cost during classification by considering higher costs for the misclassification of positive examples from the minority class as compared to the negative examples from the majority class. Such costs are coded as penalties in a cost matrix either by experts or through other heuristic approaches. Cost-sensitive learning is less practical than the other approaches, as finding the optimal cost matrix turns out to be another time-consuming problem within the class-imbalance problem (Elkan 2001).

Another possible threat concerns the construct validity of the concept of best answers. In our study, the best answer in a question thread is the one marked as accepted by the original asker. However, sometimes the best answer is not actually the one accepted – i.e., found to be useful by the original asker – but rather the one found to be the most useful by the whole community. Therefore, we acknowledge that in our study we are not predicting the '*absolute best answer*', but rather the '*fastest and good-enough answer*' that provides a prompt and effective solution to the problem reported by the asker. This choice is also consistent with the conceptualization of best answer in Stack Overflow, which we use to train the models for cross-platform prediction.

Furthermore, albeit prior work has found the presence of code snippets to be a strong indicator of the quality of answers (Asaduzzaman et al. 2013; Duijn et al. 2015), we could not include it as a predictor in our models. In fact, while in Stack Overflow the presence of code is easily verified by checking for the presence of text surround by <code> tags, this is not true for the other Q&A platforms, where code snippets mix with natural language text. As such, code snippets from the Stack Overflow were discarded during preprocessing.

Finally, we identified a couple of threats to conclusion validity. First, we acknowledge that, because of the small sample size, the Wilcoxon signed-rank test may have a reduced statistical power (Conover 1999). Second, regarding the reduced size of the datasets retrieved from the legacy forums, we acknowledge that this may have influenced our findings and conclusions. However, none of the existing legacy developer forums has ever been as successful as Stack Overflow is. As such, even with larger legacy forums, their datasets would differ in size from Stack Overflow as well.

## 10 Conclusion

In this paper, we analyzed the problem of predicting best answers in Q&A sites used by software engineers and investigate the specific challenges it poses due to its nature of class imbalance problem. We evaluated 26 best-answer prediction models through a two-stage study. In the first stage of within-platform prediction, after automated parameter tuning, we found gbm and xgbTree to build the best models for best-answer prediction in Stack Overflow. Besides, in the second stage of the study, we made – to the very best of our knowledge – the first attempt to perform cross-platform best-answer prediction. We found that best-answer prediction models trained on Stack Overflow can predict best answers consistently across both modern and legacy Q&A platforms; in addition, their performance is comparable to that achieved by models built and tested on the four test sets in a within-platform prediction setting. Overall, these findings suggest that our approach to cross-platform best-answer prediction is generalizable across sites and robust.

From a research perspective, this work provides an initial benchmark for further best-answer prediction studies, with recommended measures and data from five different datasets made publicly available.

From a practical perspective, our research has the potential to ensure the quality of the content hosted on Q&A sites and help with the problem of unresolved questions. Hence, we provide Q&A platform designers with new insights upon which to build automatic, best-answer selection tools that may help information seekers find good answers by also relying on features other than the number of answer upvotes. Besides, such features can be computed on the fly by tools since they are computationally inexpensive and need to be updated only when a new answer is added to the question thread.

Our research can also help mitigate the risk of knowledge loss as legacy support sites by supporting the automatic migration towards modern platforms. Although to date none of the available modern Q&A platforms allows to import existing content from other sources (e.g., question posting will be available in Stack Overflow API v3), even when technically feasible, the actual migration of content will pose further research challenges that we intend to face, such as coping with different interaction styles, ensuring quality of imported content, and dealing with user reputation.

As future work, we also want to improve the performance of models for cross-platform best-answer prediction. In particular, we want to investigate the effects of using mixed (i.e., within- and cross-) platform data to build the training set as well as to explore the possibility of using more computationally-intensive semantic features to leverage the textual coherence between the content of a question and its answers. Besides, we intend to deepen our understanding of how the performance of prediction models is affected when performance metrics other than AUC are selected during parameter optimization. Specifically, rather than studying Stack Overflow as a whole, we intend to investigate whether the subtopics contained therein (e.g., Java,. NET, R) affect the performance of classifiers and whether there are any sub-topics that work best in the cross-platform setting. Also, we are going to enlarge the set of predicting features. In particular, we intend to employ natural language processing and sentiment analysis techniques in order to explore the predictive potential of the emotional style conveyed through text. Finally, considering the good performance achieved by our classifier when predicting best-answers within Stack Overflow, we intend to extend our approach for ranking of all the answers within a given question thread by 'quality,' with the top-ranked answer being the accepted solution.

**Acknowledgements** We thank Stack Overflow for providing their data. We also thank Burak Turhan, for his comments on cross-context defect prediction, and Margaret-Anne Storey, Alexey Zagalsky, and Daniel M. German for their feedback on the study. This work is partially supported by the project 'EmoQuest - Investigating the Role of Emotions in Online Question & Answer Sites', funded by the Italian Ministry of Education, University and Research (MIUR) under the program "Scientific Independence of young Researchers" (SIR). The computational work has been executed on the IT resources made available by two projects, ReCaS and PRISMA, funded by MIUR under the program "PON R&C 2007-2013."

**Publisher's Note** Springer Nature remains neutral with regard to jurisdictional claims in published maps and institutional affiliations.